\documentclass[%
 aip,
 amsmath,amssymb,
 reprint,%
twocolumn
]{revtex4-1}

\pdfoutput=1
\usepackage[dvips,dvipdfmx]{graphicx}
\usepackage{dcolumn}
\usepackage{bm}

\usepackage[utf8]{inputenc}
\usepackage[T1]{fontenc}
\usepackage{mathptmx}
\usepackage{etoolbox}
\usepackage{color}

\makeatletter
\def\@email#1#2{%
 \endgroup
 \patchcmd{\titleblock@produce}
  {\frontmatter@RRAPformat}
  {\frontmatter@RRAPformat{\produce@RRAP{*#1\href{mailto:#2}{#2}}}\frontmatter@RRAPformat}
  {}{}
}%
\makeatother
\begin{document}

\preprint{AIP/123-QED}
\newcommand{\diffColor}{black}

\title[Time evolutions of information entropies in a one-dimensional Vlasov-Poisson system]{Time evolutions of information entropies in a one-dimensional Vlasov-Poisson system}


\author{K. Maekaku}
 \email{maekaku.koki20@ae.k.u-tokyo.ac.jp}
 \affiliation{Graduate School of Frontier Science, The University of Tokyo, Kashiwa 277-856, Japan}

\author{H. Sugama}%
\affiliation{National Institute for Fusion Science, Toki 509-5292, Japan}%
 \affiliation{Graduate School of Frontier Science, The University of Tokyo, Kashiwa 277-856, Japan}

\author{T.-H. Watanabe}
\affiliation{Department of Physics, Nagoya University, Nagoya 464-8602, Japan}%

\date{\today}

\begin{abstract}
A one-dimensional Vlasov-Poisson system is considered to elucidate 
how the information entropies of the probability distribution functions 
of the electron position and velocity variables 
evolve in the Landau damping process. 
Considering the initial condition given by the Maxwellian velocity distribution 
with the spatial density perturbation in the form of the cosine function of the position, 
we derive linear and quasilinear analytical solutions that accurately describe 
both early and late time behaviors of the distribution function and the electric field. 
The validity of these solutions is confirmed by comparison with numerical simulations 
based on contour dynamics. 
Using the quasilinear analytical solution, 
the time evolutions of the velocity distribution function 
and its kurtosis indicating deviation from the Gaussian distribution
are evaluated with the accuracy of the squared perturbation amplitude.  
We also determine the time evolutions of the information entropies of the electron position and velocity variables and their mutual information. 
We further consider Coulomb collisions which relax 
the state in the late-time limit in the collisionless process 
to the thermal equilibrium state. 
In this collisional relaxation process, 
 the mutual information of the position and velocity variables decreases to zero 
 while the total information entropy of the phase-space distribution function 
 increases by the decrease in the mutual information and demonstrates 
 the validity of Boltzmann's H-theorem.

\end{abstract}

\maketitle
\section{Introduction}

Landau damping is one of the most intriguing physical processes involving 
collective interactions between waves and particles in collisionless plasmas.~\cite{Landau} 
Despite the time reversibility of the Vlasov-Poisson equations, 
their solution shows that, as time progresses, 
plasma oscillations are Landau-damped 
and the electric field energy is converted into the kinetic energy of particles. 
The effects of Landau damping are universally observed in various plasma wave-particle resonance phenomena, 
such as drift waves and geodesic acoustic mode (GAM) oscillations in slab and toroidal magnetic field configurations 
in space and fusion plasmas, and have been 
the subject of extensive theoretical research.~\cite{Case,VanKampen,Nicholson,
Hazeltine-Waelbroeck,Prigogine,Zocco,Betrand,FFChen,Hammett,Villani,Zonca,Sugama2006a,Sugama2006b,Smolyakov}
%
%

The most commonly used theoretical model to study Landau damping 
is the one-dimensional Vlasov-Poisson system 
which consists of electrons with ions treated as uniform background positive charge with infinite mass. 
The solution of the linear initial value problem derived by Case and Van Kampen for this system
 is well known.~\cite{Case,VanKampen,Nicholson,Hazeltine-Waelbroeck}
The behavior of the late-time solution exhibiting Landau damping 
can be well approximated by using only eigenfrequencies with the slowest damping rate. 
In this study, the effect of an infinite number of complex eigenfrequencies is included to 
represent the early-time behavior of the electric field.
\textcolor{\diffColor}{Incidentally, it is shown in Ref.~\cite{Zocco} that the plasma dispersion function can be expressed in the form of an infinite continued fraction. From this fact, we can also understand that an infinite number of complex eigenfrequencies exist as zeros of the dielectric function. 
It is also shown in Ref.~\cite{Betrand} that a very high number of poles are required for the correct calculation of density and pressure at an early stage in the Vlasov-Poisson system.}

When treating the position $x$ and velocity $v$ of electrons as random variables, denoted as $X$ and $V$ respectively, 
the information entropy~\cite{Information} 
$S_p(X,V) \equiv - \int \int p(x, v, t) \log p(x, v, t) \, dx \, dv$ derived from the joint probability density function $p(x, v, t)$ of $(X, V)$ is known to be one of 
the Casimir invariants in the Vlasov-Poisson system,~\cite{Casimir,Maekaku}
and it does not depend on time $t$. 
On the other hand, the entropies $S_p(X) \equiv - \int p_X(x, t) \log p_X(x, t) \, dx$ and $S_p(V) \equiv - \int p_V(v, t) \log p_V(v, t) \, dv$ determined from the marginal probability density functions $p_X(x, t) = \int p(x, v) \, dv$ and $p_V(v, t) = \int p(x, v) \, dx$ of $X$ and $V$ respectively, are allowed to vary with time.
In the present work, we derive novel analytical expressions that accurately represent the early-time behaviors of 
the linear solutions for the electric field and the distribution function by series expansions in time and velocity variables. 
Combining these early-time expressions with the late-time approximate solutions, 
we accurately determine the time evolution of the spatially averaged velocity distribution function of electrons 
obtained from quasilinear theory.~\cite{Drummond,Vedenov} 
Using these linear and quasilinear analytical solutions, we clarify the time evolution of 
the information entropies $S_p(X)$, $S_p(V)$, 
and the mutual information $I(X, V) \equiv S_p(X) + S_p(V) - S_p(X, V)$ associated with the Landau damping.
\textcolor{\diffColor}{We here note that, in Ref.~\cite{Cassak}, the energy conversion in phase space density moments in the Vlasov-Maxwell system is investigated by separating the kinetic entropy [which corresponds to $S_p(X,V)$] to the two parts [which correspond to $S_p(X)$ and $S_p(X,V)-S_p(X)$].
Also, in Ref.~\cite{Ghizzo}, the role of the entropy production in momentum transfer in the Vlasov-Maxwell system is discussed based on the information theory.}

We note that 
the quasilinear solution for the background velocity distribution function at time $t$ is obtained by integrating the product of the electric field and the linear solution of the perturbed distribution function over time from the initial time to $t$. 
Therefore, using only the linear solutions represented by a few complex eigenfrequencies is insufficient.
We emphasize that, to describe the time evolutions of the entropies and the mutual information, 
it is necessary to use accurate expressions of the linear solutions near the initial time 
as derived in this study. 

The validity of the obtained linear and quasilinear solutions is confirmed by comparison with simulation results 
based on contour dynamics.~\cite{Zabusky,Sato}
We also obtain the velocity distribution $p_V$ in the limit of $t \rightarrow +\infty$ and show how 
it deviates from the Gaussian distribution. 
In addition, we consider the effects of Coulomb collisions, which relax the distribution function to the thermal equilibrium state, 
decrease the mutual information $I(X, V, t)$ to zero, increase the entropy $S_p(X,V)$, 
and thus validate Boltzmann's H-theorem.~\cite{H-theorem}

The rest of this paper is organized as follows.
In Sec.~II, the basic equations of the Vlasov-Poisson system are presented, 
and the solution to its linear initial value problem is provided using the Laplace transform. 
The validity of the linear analytical solutions for the electric field and the perturbed distribution function is confirmed 
by comparison with contour dynamics simulations with small initial perturbation amplitudes.
In Sec.~III, using approximate expressions for complex eigenfrequencies with large absolute values, 
an approximate integral formula for the electric field incorporating the effects of an infinite number of complex eigenfrequencies is derived. 
In addition, the linear analytical solutions for the electric field and the distribution function near the initial time 
are expressed as series expansions in time and velocity variables.
In Sec.~IV, using the results of Secs.~II and III, 
a quasilinear analytical solution describing the time evolution of the background velocity distribution function of electrons is derived, 
and its validity is confirmed by contour dynamics simulations as well. 
By using this quasilinear analytical solution, the time evolution of the electron kinetic energy increases due to Landau damping, 
the velocity distribution function in the limit of $t \rightarrow +\infty$, and 
physical quantities, such as the kurtosis representing a deviation from the Gaussian distribution, are accurately determined.
In Sec.~V, the time evolution of the information entropies of the position and velocity variables of electrons and 
the mutual information is determined with an accuracy of the order of the squared perturbation amplitude.
In Sec.~VI, the changes in the entropies and the mutual information from the collisionless process 
to the thermal equilibrium state due to Coulomb collisions are evaluated.
Finally, conclusions and discussion are given in Sec.~VII.

\section{Linear Analysis of Vlasov-Poisson system}

\subsection{Vlasov-Poisson system}

Under the assumption that there is no magnetic field,  
the Vlasov equation for electrons is given by 
\begin{equation}
\label{Vlasov0}
\frac{\partial f({\bf x}, {\bf v}, t)}{\partial t}
+
{\bf v} \cdot 
\frac{\partial f({\bf x}, {\bf v}, t)}{\partial {\bf x}}
- 
\frac{e}{m} {\bf E}({\bf x}, t) \cdot
\frac{\partial f({\bf x}, {\bf v}, t)}{\partial {\bf v}}
=
0
,
\end{equation}
where $-e$ and $m$ are the electron charge and mass, respectively, 
and $E(x, t)$ represents the electric field. 
The distribution function $f({\bf x}, {\bf v}, t)$ of electrons
is defined such that 
$f({\bf x}, {\bf v}, t) d^3 x d^3 v$ represents the number of electrons in the phase space volume element 
$d^3x \; d^3v$ around $({\bf x}, {\bf v})$ at time $t$. 
We ignore the motion of ions that have the uniform density $n_0$ and infinite mass. 
Then, the electric field ${\bf E}({\bf x}, t)$ is determined by Poisson's equation, 
\begin{equation}
\label{Poisson0}
\nabla \cdot {\bf E}({\bf x}, t)
= 4 \pi e
\left(
n_0 
- \int d^3v \; f({\bf x}, {\bf v}, t)
\right)
.
\end{equation}

We now assume that $f({\bf x}, {\bf v}, t)$ and ${\bf E}({\bf x}, t)$ are independent of $y$ and $z$, and that $E_y = E_z =0$. 
Integration of $f({\bf x}, {\bf v}, t)$ with respect to $v_y$ and $v_z$ is done to define
\begin{equation}
f (x, v_x, t)
\equiv
\int_{-\infty}^{+\infty} dv_y \int_{-\infty}^{+\infty} dv_z \; 
f({\bf x}, {\bf v}, t)
.
\end{equation}
Then, from Eq.~(\ref{Vlasov0}), we obtain the Vlasov equation 
in the $(x, v)$ phase space space 
as 
\begin{equation}
\label{Vlasov1}
\frac{\partial f(x, v, t)}{\partial t}
+
v
\frac{\partial f(x, v, t)}{\partial x}
- 
\frac{e}{m} E(x, t) 
\frac{\partial f(x, v, t)}{\partial v}
=
0
,
\end{equation}
where $v = v_x$ and $E(x, t) = E_x (x, t)$ are used. 
Poisson's equation in Eq.~(\ref{Poisson0}) is rewritten as 
\begin{equation}
\label{Poisson1}
\frac{\partial E(x, t)}{\partial x}
= 4 \pi e
\left(
n_0 
- \int_{-\infty}^{+\infty} d v \; f(x, v, t)
\right)
.
\end{equation}
Hereafter, we consider the structure of the distribution function $f(x, v, t)$ on the two-dimensional $(x, v)$ phase space instead of the six-dimensional $({\bf x}, {\bf v})$ space. 
The number of electrons in the area element $dx \; dv$ about the point $(x, v)$ in the phase space 
at time $t$ is given by $f(x, v, t) dx \; dv$.

\subsection{Linearization}

We now write 
the distribution function $f(x, v, t)$ as 
the sum of the equilibrium and perturbation parts, 
\begin{equation}
\label{f0f1}
f(x, v, t)
=
f_0 (v) + f_1 (x, v, t)
.
\end{equation}
Here the equilibrium distribution function is given by $f_0(v)$ which satisfies
\begin{equation}
\label{f0n0}
\int d^3 v \; f_0(v) = n_0
.
\end{equation}
The perturbed distribution function $ f_1 (x, v, t)$ and the 
electric field $E (x, t)$ are assumed to be given by 
\begin{eqnarray}
\label{f1E}
f_1 (x, v, t)  & =  & \mbox{Re}[ f_1(k, v, t) \exp ( i \; k \;  x ) ]
,
\nonumber \\ 
E (x, t) & = & \mbox{Re}
[ E (k, t)  \exp ( i \; k \;  x )  ]
,
\end{eqnarray}
where the wavenumber in the $x$-direction is given by $k > 0$. 
Substituting Eq.~(\ref{f0f1}) into Eq.~(\ref{Vlasov1}) and using Eq.~(\ref{f1E}), 
we obtain the linearized Vlasov equation, 
\begin{equation}
\label{Vlasov2}
\frac{\partial f_1(k, v, t) }{\partial t}
+ i \; k \; v \; 
f_1(k, v, t) 
- 
\frac{e}{m} E(k, t) \frac{\partial f_0 (v)}{\partial v}
=
0
,
\end{equation}
where the nonlinear term 
\begin{equation}
- \frac{e}{m} E (x, t) 
\frac{\partial  f_1 (x, v, t)}{\partial v}
\end{equation}
is neglected as a small term. 
Using Eqs.~(\ref{f0f1}) and (\ref{f0n0}), 
Poisson's equation in Eq.~(\ref{Poisson1}) is rewritten as 
\begin{equation}
\label{Poisson2}
i \: k \;  E(k,t)
=
- 4 \pi e
\int_{-\infty}^{+\infty} d v \; f_1(k, v, t) 
.
\end{equation}

\subsection{Laplace transform}

We now use the Laplace transform to solve the linearized Vlasov-Poisson equations 
given by Eq.~(\ref{Vlasov2}) and (\ref{Poisson2}). 
The Laplace transforms of $f_1(k, v, t)$ and $E(k, t)$ are denoted by 
$
f_1(k, v, \omega)
$
and 
$
E(k, \omega)
$,
respectively. 
\textcolor{\diffColor}{
Instead of the variable \( p \) used in the conventional Laplace transform, we put \( p = -i \omega \) and employ the complex frequency \( \omega \) to perform the analysis in the manner similar to the Fourier transform as seen in Ref.~\cite{Nicholson}.}
The inverse Laplace transform gives 
\begin{equation}
f_1(k, v, t)
= \int_{L_f} \frac{d\omega}{2\pi} f_1 (k, v, \omega) e^{-i \omega t}
,
\end{equation}
and 
\begin{equation}
E(k, t)
= \int_{L_E} \frac{d\omega}{2\pi}E (k,\omega) e^{-i \omega t}
,
\end{equation}
where the Laplace contours $L_f$ and $L_E$ need to pass above all 
poles of $f_1 (k, v, \omega)$ and $E (k,\omega)$ on the complex $\omega$-plane, 
respectively. 
Now, the Vlasov equation and Poisson's equation in Eqs.~(\ref{Vlasov2}) and (\ref{Poisson2}) 
are represented in terms of $f_1 (k, v, \omega)$ and $E (k,\omega)$ by  
\begin{equation}
\label{Vlasov3}
( - i  \; \omega + i \; k \; v ) f_1(k, v, \omega)
-
\frac{e}{m} E(k, \omega) \frac{\partial f_0 (v)}{\partial v}
= f_1 (k, v, t = 0)
,
\end{equation}
and
\begin{equation}
\label{Poisson3}
i \; k \; E(k, \omega) 
=
- 4 \pi e
\int_{-\infty}^{+ \infty} dv \; f_1(k, v, \omega)
,
\end{equation}
respectively. 

Solving Eqs.~(\ref{Vlasov3}) and (\ref{Poisson3}) for 
the perturbed distribution function $f_1(k, v, \omega)$ and 
the electric field $E(k, \omega)$, 
we obtain 
\begin{equation}
f_1(k, v, \omega)
=
\frac{
(e/m) E(k, \omega) \partial f_0 (v)/ \partial v
+ f_1 (k, v, t = 0)
}{- i  \; \omega + i \; k \; v }
,
\end{equation}
and 
\begin{equation}
E(k, \omega)
=
\frac{
4 \pi e
}{k^2 \epsilon (k, \omega)}
\int_{-\infty}^{+\infty} dv \;
\frac{f_1 (k, v, t = 0)
}{ v - \omega / k }
,
\end{equation}
where the dielectric function $\epsilon (k, \omega)$ is defined by 
\begin{equation}
\label{dielectric0}
\epsilon (k, \omega)
=
1 - 
\frac{
\omega_p^2
}{n_0 k^2}
\int_{-\infty}^{+\infty} dv \;
\frac{\partial f_0 (v)/\partial v
}{ v - \omega / k }
.
\end{equation}
Here, the plasma frequency is defined by 
$\omega_p \equiv (4 \pi n_0 e^2/ m)^{1/2}$. 
Now,  we define 
\begin{equation}
\label{Fkomega}
F(k, \omega)
\equiv
\int_{-\infty}^{+\infty} dv \;
\frac{f_1 (k, v, t = 0)}{ v - \omega / k }
,
\end{equation}
to write 
$E(k, t)$ 
and 
$f_1(k, v, t)$ 
for $t > 0$
as 
\begin{equation}
E(k, t)
= 
\frac{4 \pi e}{k^2}
\int_{L_E} \frac{d\omega}{2\pi}  e^{-i \omega t}
\frac{F(k, \omega)
}{\epsilon (k, \omega)}
,
\end{equation}
and 
\begin{eqnarray}
f_1(k, v, t)
&=&\int_{L_f} \frac{d\omega}{2\pi}  
\frac{e^{-i \omega t}}{- i  (\omega  - k \, v ) }\nonumber\times\\&&
\left[
\frac{\omega_p^2}{n_0 k^2}
\frac{F(k, \omega)}{\epsilon(k, \omega)}
\frac{ \partial f_0 (v)}{\partial v}
+ f_1 (k, v, t = 0)
\right]
,
\end{eqnarray}
respectively. 
Let us define complex-valued eigenfrequencies 
$\{ \omega_\mu \}$ as
zeros of $\epsilon(k, \omega)$, 
\begin{equation}
\epsilon(k, \omega_\mu)
= 0\label{epsilon_is_zero}
.
\end{equation}
Then, we can write 
\begin{equation}
\label{Ekt}
E(k, t)
= 
- i 
\frac{4 \pi e}{k^2}
\sum_\mu
e^{-i \omega_\mu t}
\frac{F(k, \omega_\mu)
}{\partial_\omega\epsilon (k, \omega_\mu)}
,
\end{equation}
and 
\begin{eqnarray}
\label{f1kvt}
f_1(k, v, t)
& = & 
e^{-i k \, v \, t}
\left[
f_1 (k, v, t = 0)
+ 
\frac{\omega_p^2}{n_0 k^2}
\frac{F(k, k \, v)}{\epsilon(k, k \, v)}
\frac{ \partial f_0 (v)}{\partial v}
\right]
\nonumber 
\\ & & \mbox{}
+
\frac{\omega_p^2}{n_0 k^2}
\frac{ \partial f_0 (v)}{\partial v}
\sum_\mu 
\frac{e^{-i \omega_\mu t}}{
(\omega_\mu - k \, v)}
\frac{F(k, \omega_\mu)
}{\partial_\omega\epsilon (k, \omega_\mu)}
,
\end{eqnarray}
where the derivative with respect to the complex-valued frequency $\omega$ 
is represented by 
$\partial_\omega \equiv \partial/ \partial \omega$.

\subsection{Conditions of equilibrium and initial perturbation}

Hereafter,  we assume that the equilibrium function $f_0(v)$ is given by the Maxwellian 
\begin{equation}
\label{Maxwellian}
f_0 (v)
= 
n_0 \sqrt{
\frac{m}{2\pi T}
}
\exp
\left( 
- \frac{m v^2}{2 T}
\right)
= 
\frac{n_0}{\sqrt{\pi} v_T}
\exp
\left( 
- \frac{v^2}{v_T^2}
\right)
, 
\end{equation}
where 
$T$ represents the equilibrium temperature and 
\begin{equation}
v_T \equiv \sqrt{2} v_t \equiv 
\sqrt{2T/m}
\end{equation}
is used. 
Then, the dielectric function $\epsilon (k, \omega)$ in Eq.~(\ref{dielectric0}) 
is expressed by 
\begin{equation}
\label{dielectric1}
\epsilon (k, \omega)
=
1 + 
\frac{1}{k^2 \lambda_D^2}
\left[ 1 + \frac{\omega}{k v_T}
Z \left( \frac{\omega}{k v_T} \right)
\right]
,
\end{equation}
where the Debye length 
$\lambda_D$ is defined by 
\begin{equation}
\lambda_D
\equiv \frac{v_t}{\omega_p}
\equiv \sqrt{\frac{T}{4\pi n_0 e^2}}
\end{equation}
and the plasma dispersion function $Z(\zeta)$ 
is defined by Eq.~(\ref{Z}) in Appendix~A. 

Using Eq.~(\ref{dielectric1}), 
the dispersion relation in Eq.~(\ref{epsilon_is_zero}) is 
rewritten as 
\begin{equation}
\label{dispersion_relation}
\epsilon (k, \omega_\mu)
=
1 + 
\kappa^{-2}
\left[ 1 + \zeta_\mu
Z ( \zeta_\mu )
\right]
= 
0
,
\end{equation}
where $\zeta_\mu = \omega_\mu / k v_T$ and $\kappa \equiv k \lambda_D$. 
Figure 1 shows the distribution of complex eigenfrequencies \(\omega_{\mu} = \omega_{\mu r} + i \gamma_{\mu}\) in the complex plane, 
calculated from Eq.~(\ref{dispersion_relation}) for \(k\lambda_D = 1/2\). 
There are an infinite number of complex eigenfrequencies.
The complex eigenfrequency with $\omega_r >0$ and the smallest $|\gamma|$ 
is given by  $\omega_{\mbox{\o}} = ( 1.41566 - 0.153359 i ) \omega_p$. 
From Eqs.~(\ref{dispersion_relation}) and (\ref{Zcc}), we find 
\begin{equation}
[ \epsilon (k, - \omega^*) ]^*
=
\epsilon (k, \omega)
,
\end{equation}
and 
\begin{equation}
 \epsilon (k, \omega_\mu) = 0
\hspace*{5mm}
\Rightarrow
\hspace*{5mm}
 \epsilon (k, - \omega_\mu^*) = 0
 ,
\end{equation}
where $^*$ represents the complex conjugate. 
Thus, if $\omega_\mu = \omega_{\mu r} + i \gamma_{\mu}$ is a zero of the dielectric function, 
$- \omega_\mu^*=  -\omega_{\mu r} + i \gamma_{\mu}$ is so, too. 
Therefore, 
complex eigenfrequencies are distributed in pairs symmetric with respect to the \({\rm Im} (\omega)\)-axis
as seen in Fig.~1. 

\begin{figure}[h]
\centering
\includegraphics[width=8cm]{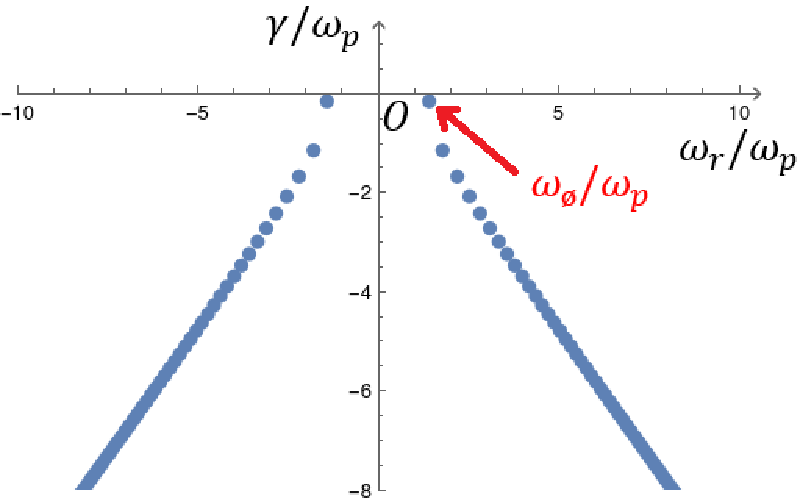}
\caption{Complex eigenvalues of frequency $\omega = \omega_r+i\gamma$ ($k\lambda_D=1/2$). The horizontal and vertical axes represent the real part $\omega_r$ and the imaginary part (or the growth rate) $\gamma$, respectively. 
The complex eigenfrequency with $\omega_r >0$ and the smallest $|\gamma|$ is given by  
$\omega_{\mbox{\o}} = ( 1.41566 - 0.153359 i ) \omega_p$. 
There are an infinite number of eigenfrequencies other than $\omega_{\mbox{\o}}$.
}
\label{fig:omegaAlpha1}
\end{figure}

The derivative of $\epsilon (k, \omega)$ with respect to $\omega$ is 
given by 
\begin{equation}
\label{dedomega}
\frac{\partial \epsilon (k, \omega)}{\partial \omega}
=
\frac{1}{k v_T}
\frac{\partial \epsilon}{\partial \zeta}
= 
\frac{1}{k v_T} \frac{1}{k^2 \lambda_D^2}
\left[
- 2 \zeta
+ (1 - 2 \zeta^2)
Z(\zeta)
\right]
\end{equation}
where Eqs.~(\ref{dielectric1}), (\ref{dZ}), and $\zeta = \omega / k v_T$ are used. 
Then, from Eq.~(\ref{dispersion_relation}) and (\ref{dedomega}), we find
\begin{eqnarray*}
\label{dedomega2}
\partial_\omega \epsilon (k, \omega_\mu)
& = & 
\frac{1}{\omega_\mu}
\left[
- \left( 1 +  \frac{1}{k^2 \lambda_D^2} \right)
+ 2 \zeta_\mu^2
\right]
\\ & = & 
\frac{1}{k^2 \lambda_D^2\omega_\mu}
\left[
- 1 - k^2 \lambda_D^2 
+ \frac{\omega_\mu^2}{\omega_p^2}
\right]
\end{eqnarray*}

We now impose the initial condition given by 
\begin{equation}
\label{f1t0}
f_1(k, v, t = 0) = \alpha f_0 (v)
,
\end{equation}
where $f_0(v)$ is the Maxwellian equilibrium distribution function given by Eq.~(\ref{Maxwellian}) 
and 
$\alpha$ is a small constant. 
Then, from Poisson's equation, we have 
\begin{equation}
E(k, t = 0) = i \frac{4\pi n_0 e }{k}\alpha  
= i \frac{m}{e} \frac{\omega_p}{k} \alpha
\end{equation}
Using the initial condition in Eq.~(\ref{f1t0}),  
the plasma dispersion function in Eq.~(\ref{Z}), 
and Eq.~(\ref{Fkomega}), 
we obtain
\begin{equation}
F(k, \omega)
\equiv
\int_{-\infty}^{+\infty} dv \;
\frac{f_1 (k, v, t = 0)}{ v - \omega / k }
=
\alpha \frac{n_0}{v_T} 
Z \left( \frac{\omega}{k v_T} \right)
\label{solFko}
,
\end{equation}
and 
\begin{equation}
F(k, k v)
\equiv
\int_{-\infty}^{+\infty} dv' \;
\frac{f_1 (k, v', t = 0)}{ v' - v }
=
\alpha \frac{n_0}{v_T} 
Z \left( \frac{v}{v_T} \right)
\label{solFkv}
,
\end{equation}
which are used to evaluate $E(k, t)$ and $f_1(k, v,  t)$ 
given in Eqs.~(\ref{Ekt}) and (\ref{f1kvt}), respectively.

\subsection{Solution of initial value problem}

We now substitute Eqs.~(\ref{dedomega2}), (\ref{solFko}), 
and (\ref{solFkv}) 
into Eqs.~(\ref{Ekt}) and (\ref{f1kvt})
to express the electric field and the perturbed distribution function 
at time $t > 0$ as 
\begin{eqnarray}
E(k, t)\nonumber
& = & 
\frac{4 \pi e}{k^2}
\int_{L_E} \frac{d\omega}{2\pi}  e^{-i \omega t}
\frac{F(k, \omega)
}{\epsilon (k, \omega)}\\\nonumber
&=&
\alpha
\frac{4 \pi e n_0}{k}
\int_{-\infty}^{+\infty} \frac{d \zeta}{2\pi} \;
\frac{Z(\zeta)e^{-i \zeta \tau}}{1 + \kappa^{-2}
[ 1 + \zeta Z(\zeta) ]}
\\\nonumber & = & 
- i 
\frac{4 \pi e}{k^2}
\sum_\mu
e^{-i \omega_\mu t}
\frac{F(k, \omega_\mu)
}{\partial_\omega\epsilon (k, \omega_\mu)}
\\ \nonumber
& = & 
- i \; \alpha
\frac{m \omega_p^2}{e k} 
k^2 \lambda_D^2 ( 1+ k^2 \lambda_D^2 )\nonumber \\ & &
\times \sum_\mu
\frac{e^{-i \omega_\mu t}
}{1 +  k^2 \lambda_D^2 - \omega_\mu^2/\omega_p^2}\nonumber
\\ 
& = & 
- i \; \alpha
\frac{m \omega_p^2}{e k} 
k^2 \lambda_D^2 ( 1+ k^2 \lambda_D^2 )\nonumber
\\&&
\times \sum_{\mu, \; {\rm Re} \omega_\mu > 0}
2 {\rm Re} \left[
\frac{e^{-i \omega_\mu t}
}{1 +  k^2 \lambda_D^2 - \omega_\mu^2/\omega_p^2}
\right]
,
\label{solE}
\end{eqnarray}
and 
\begin{eqnarray}
& & 
\hspace*{-5mm}
f_1(k, v, t)
=\nonumber
\\&&\int_{-\infty}^{+\infty} \frac{d\omega}{2\pi}  
\frac{e^{-i \omega t}}{- i  (\omega  - k \, v ) }
\Biggl[
f_1 (k, v, t = 0)\nonumber \\&&+
\frac{\omega_p^2}{n_0 k^2}
\frac{F(k, \omega)}{\epsilon(k, \omega)}
\frac{ \partial f_0 (v)}{\partial v}
\Biggr]\nonumber
\\
& & 
\hspace*{-5mm}
=
\alpha f_0 (v) 
\Biggl[ 
\exp \left( - i \frac{v}{v_T} \tau \right)\nonumber
\\&-& i \kappa^{-2}\frac{v}{v_T}
\int_{-\infty}^{+\infty} \frac{d\zeta}{2\pi}  
\; 
\frac{Z(\zeta)e^{-i \zeta \tau}}{(\zeta - v / v_T ) \{ 1 + \kappa^{-2}
[ 1 + \zeta Z(\zeta) ]\} }
\Biggr]\nonumber
\\
& & 
\hspace*{-5mm}
=
e^{-i k \, v \, t}
\Biggl[
f_1 (k, v, t = 0)
+ 
\frac{\omega_p^2}{n_0 k^2}
\frac{F(k, k \, v)}{\epsilon(k, k \, v)}
\frac{ \partial f_0 (v)}{\partial v}
\Biggr]\nonumber
\\&+& 
\frac{\omega_p^2}{n_0 k^2}
\frac{ \partial f_0 (v)}{\partial v}
\sum_\mu 
\frac{e^{-i \omega_\mu t}}{
(\omega_\mu - k \, v)}
\frac{F(k, \omega_\mu)
}{\partial_\omega\epsilon (k, \omega_\mu)}\nonumber
\\
&  &  
\hspace*{-5mm}
=
\alpha f_0 (v) 
\Bigg[ 
e^{-i k \, v \, t}
\left\{ 1  - 
\frac{(v/v_T) Z(v/v_T)}{k^2 \lambda_D^2 + [ 1 + (v/v_T) Z(v/v_T) ] }
\right\}\nonumber
\\&-& k \; v \; 
( 1 + k^2 \lambda_D^2) 
\sum_\mu 
\frac{e^{-i \omega_\mu t}}{
(\omega_\mu - k \, v)
(1 +  k^2 \lambda_D^2 - \omega_\mu^2/\omega_p^2)}
\Bigg]
,
\label{solf1}
\end{eqnarray}
respectively, where the normalized time $\tau \equiv k v_T t$ and
the residue theorem are used.  
\textcolor{\diffColor}{
Appendix B presents supplementary explanations about contours used for the integrals in Eqs.~(\ref{solE}) and (\ref{solf1}).
}

\begin{figure}[h]
\centering
\includegraphics[width=8cm]{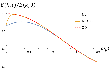}
\caption{Time evolution of $E(k,t)/E(k,0)$ for $k\lambda_D = 1/2$. 
The red dashed curve represents the result from the CD simulation using $\alpha = 0.01$. 
The solid curves are obtained from Eq.~(\ref{solE}) where the summation over $\mu$ 
is done by including a finite number of pairs of the complex eigenfrequencies 
$(\omega_\mu, -\omega_\mu^*)$ from 
$(\omega_{\mbox{\o}}, -\omega_{\mbox{\o}}^*)$ to the pair 
with the $N$th smallest decay rate $|\gamma_\mu|$.  
Here, the case for $N=2$ and 100 are shown.
}
\label{fig:Et1000}
\end{figure}

Figure~\ref{fig:Et1000} shows \(E(k,t)\) calculated as a function of time using Eq.~(\ref{solE}) for \(k\lambda_D = 1/2\). 
The summation over \(\mu \) in Eq. (\ref{solE}) is done 
by including a finite number of pairs of the complex eigenfrequencies \((\omega_{\mu}, -\omega_{\mu}^*)\) 
from $(\omega_{\mbox{\o}}, -\omega_{\mbox{\o}}^*)$ to the pair with the $N$th smallest decay rate $|\gamma_\mu|$. 
The results obtained for the cases of  \(N = 2\) and $N=100$ are shown in Fig.~2 
where the result from the contour dynamics (CD) simulation for the case of 
$\alpha = 0.01$ are 
also plotted by the red dashed curve.
The CD method and the simulation conditions used in this paper 
are explained in Appendix~C. 
A good agreement among the three results is confirmed except near \(t = 0\). 
When using the larger number $N$ of the pairs of complex eigenfrequencies, 
the better agreement with the CD simulation result is confirmed. 
Near \(t = 0\), 
even in the large \(N\) limit, 
the sum over \(\mu \) does not converge uniformly, 
and the number of pairs \(N\) required for a good convergence increases to infinity 
as \(t\) approaches 0. 
In Sec.~III, we derive an approximate expression for \(E(k,t)\) near \(t = 0\) 
that includes the effect of an infinite number of complex eigenfrequencies \(\{ \omega_{\mu} \} \).

\begin{figure*}[ht]
\centering
\includegraphics[width=18cm]{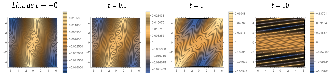}
\includegraphics[width=18cm]{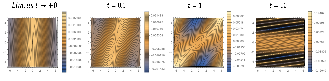}
\includegraphics[width=18cm]{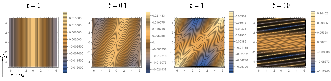}
\caption{
Contours of $f_1(x, v, t)/( \alpha^2 f_0(v) )$ in the $(x, v)$-space 
for the case of $k \lambda_D=1/2$. 
The results obtained from Eq. (\ref{solf1}) 
using 2 pairs (upper row) and 100 pairs (middle row) of complex eigenfrequencies 
and those from the CD simulation using $\alpha = 0.01$  (lower row for \(t > 0\)) 
are shown. 
}
\label{fig:f1}
\end{figure*}

Similar to the case of \(E(k,t)\), 
the effect of a larger number of the complex eigenfrequencies \(\omega_{\mu}\) 
on \( f_1(k, v, t) \) 
becomes more significant near $t = 0$. 
To accurately evaluate 
\( f_1(k, v, t) \)  from Eq. (\ref{solf1}),  
we need to include a larger number $N$ of 
complex eigenfrequency pairs as \(t \rightarrow +0\). 
However, we should note that 
the denominator of each term in the sum over \(\mu \) in the analytical solution for \(E(k,t)\) [Eq. (\ref{solE})] is 
a quadratic function of \(\omega_{\mu}\) 
while in the analytical solution for \( f_1(k, v, t) \)  [Eq. (\ref{solf1})], 
it is a cubic function of \(\omega_{\mu}\). 
Therefore, \( f_1(k, v, t) \)  
converges faster than \(E(k,t)\) as 
the number $N$ of included eigenfrequency pairs \(\omega_{\mu}\) increases, and 
\( f_1(k, v, t) \) shows a uniform convergence even near \(t = 0\).
Figure~\ref{fig:f1} shows contour plots of 
$f_1(x, v, t)/( \alpha^2 f_0(v) )$ in the \((v,t)\)-plane for \(\omega_p t = 0, 0.1, 1, 10\). 
For \(t > 0\), the plots show 
$f_1(x, v, t)/( \alpha^2 f_0(v) )$
calculated using 2 pairs (upper row) and 100 pairs (middle row) of complex eigenfrequencies 
from Eq. (\ref{solf1}) and 
$f_1(x, v, t)/( \alpha^2 f_0(v) )$ from CD simulation results for $\alpha = 0.01$ (lower row). 
The results using 2 pairs of complex eigenfrequencies 
agree with the results of CD simulation for \(\omega_p t \geq 1\). 
The results 
using 1000 pairs 
show a better agreement with CD simulation results for all \(t > 0\).

\section{Approximate representation of linear solutions in early time}

\subsection{Approximate expression for complex eigenfrequencies with large absolute values}

We here recall that the complex-valued eigenfrequency $\omega_\mu$ is determined by 
Eq.~(\ref{dispersion_relation}) using the plasma dispersion function $Z ( \zeta )$. 
When $|\zeta | \gg 1$, $|{\rm Re} \zeta | > |{\rm Im} \zeta | $, and ${\rm Im} \zeta < 0$, 
we have
\begin{equation}
\label{large_zeta}
 1 + \zeta
Z ( \zeta )
=
i \sqrt{\pi} \zeta e^{-\zeta^2}
+ {\cal O}(\zeta^{-2})
.
\end{equation}
Therefore, it is found from Eqs.~(\ref{dispersion_relation}) and (\ref{large_zeta})  
that, when 
$|\zeta_\mu | \gg 1$ and ${\rm Im} \zeta_\mu < 0$, 
$\zeta_\mu$ satisfies
\begin{equation}
\label{relation_large_zeta}
i \sqrt{\pi} \zeta_\mu e^{ -\zeta_\mu^2} = - \kappa^2 + {\cal O}(\zeta_\mu^{-2})
. 
\end{equation}
Using Eq. ~(\ref{relation_large_zeta}) and 
$
\zeta_\mu \equiv |\zeta_\mu | e^{i \theta_\mu}
$, 
we obtain
\begin{equation}
\zeta_\mu  
\simeq 
 i \frac{\kappa^2}{\sqrt{\pi}} 
e^{\zeta_\mu^2} = 
 i \frac{\kappa^2}{\sqrt{\pi}} 
\exp [ |\zeta_\mu|^2 ( \cos 2\theta_\mu + i \sin 2\theta_\mu) ]
,
\end{equation}
from which we get 
\begin{equation}
| \zeta_\mu  |
\simeq 
 \frac{\kappa^2}{\sqrt{\pi}} 
\exp ( |\zeta_\mu|^2  \cos 2\theta_\mu ) 
,
\end{equation}
and 
\begin{equation}
e^{i\theta_\mu}
\simeq 
i 
\exp ( i |\zeta_\mu|^2  \sin 2\theta_\mu ) 
.
\end{equation}
Therefore,  $\theta_\mu$ satisfies 
\begin{equation}
\label{theta_alpha}
\theta_\mu
\simeq 
\frac{\pi}{2} 
+ |\zeta_\mu|^2  \sin 2\theta_\mu  + 2 \pi n
,
\end{equation}
where $n$ is an integer. 
Then, we define $\delta_\mu$ by 
\begin{equation}
\label{delta_alpha}
2 \theta_\mu
=
- \frac{\pi}{2} 
+\delta_\mu
\end{equation}
to have 
\begin{equation}
| \zeta_\mu  |
\simeq 
 \frac{\kappa^2}{\sqrt{\pi}} 
\exp ( |\zeta_\mu|^2 \sin \delta_\mu) 
,
\end{equation}
and 
\begin{equation}
\log \left(
 \frac{\sqrt{\pi}}{\kappa^2}| \zeta_\mu  |
\right)
\simeq 
 |\zeta_\mu|^2 \sin \delta_\mu
\simeq 
 |\zeta_\mu|^2 \delta_\mu
 ,
\end{equation}
which leads to 
\begin{equation}
\delta_\mu
\simeq 
\frac{1}{ |\zeta_\mu|^2}
\log \left(
 \frac{\sqrt{\pi}}{\kappa^2}| \zeta_\mu  |
\right)
,
\end{equation}
where 
$0 < \delta_\mu \ll 1$
is assumed. 
Substituting Eq.~(\ref{delta_alpha}) into Eq.~(\ref{theta_alpha}), 
we have 
\begin{eqnarray}
- \frac{\pi}{4} + \frac{\delta_\mu}{2}
&\simeq &
\frac{\pi}{2} 
-  |\zeta_\mu|^2  \cos \delta_\mu  + 2 \pi n\nonumber\\&
\simeq&
\frac{\pi}{2} 
-  |\zeta_\mu|^2  
\left( 1 - \frac{\delta_\mu^2}{2} \right)  + 2 \pi n
,
\end{eqnarray}
which yields
\begin{equation}
 |\zeta_\mu|^2
\simeq  
 \frac{3\pi}{4} - \frac{\delta_\mu}{2}
+ 2 \pi n
.
\end{equation}

Thus, 
the approximate value of 
$
\zeta_\mu \equiv |\zeta_\mu | e^{i \theta}
$
for  $|\zeta_\mu | \gg 1$ is given by 
\begin{equation}
 |\zeta_\mu |
\simeq  
\sqrt{ 
\left(
 2  n + 
 \frac{3}{4} 
\right) \pi
}
\end{equation}
with 
\begin{equation}
\theta_\mu =
- \frac{\pi}{4} + \frac{\delta_\mu}{2}
\simeq 
- \frac{\pi}{4}
+
\frac{1}{2  |\zeta_\mu|^2}
\log \left(
 \frac{\sqrt{\pi}}{\kappa^2}| \zeta_\mu  |
\right)
\end{equation}
where the integer $n$ needs to be large, $n \gg 1$, to satisfy the condition 
$|\zeta_\mu | \gg 1$. 
Recall that when $\omega_\mu$ is a complex-valued eigenfrequency, 
$-\omega_\mu^*$ is so, too. 
Then, the approximate values of complex-valued eigenfrequency with large absolute values
are denoted by $\omega_n$ and $- \omega_n^*$ which are given by 
\begin{equation}
\label{zeta_n}
\frac{\omega_n}{k v_T} \equiv \zeta_n
 \equiv 
\sqrt{ 
\left(
 2  n + 
 \frac{3}{4} 
\right) \pi
}
\; \;  e^{i \theta_n}
,
\end{equation}
\begin{equation}
\frac{- \omega_n^*}{k v_T} \equiv 
- \zeta_n^*
\equiv 
\sqrt{ 
\left(
 2  n + 
 \frac{3}{4} 
\right) \pi
}
\; \;  e^{i (\pi - \theta_n)}\label{zeta1}
\end{equation}
and
\begin{equation}
\theta_n \equiv
- \frac{\pi}{4}
+
\frac{1}{2  |\zeta_n|^2}
\log \left(
 \frac{\sqrt{\pi}}{\kappa^2}| \zeta_n |
\right)\label{zeta2}
\end{equation}
where the integer $n$ is used instead of $\mu$ 
to specify different approximate eigenfrequencies.  
The approximation is good for $n \gg 1$.  

\begin{figure}[h]
\centering
\includegraphics[width=8cm]{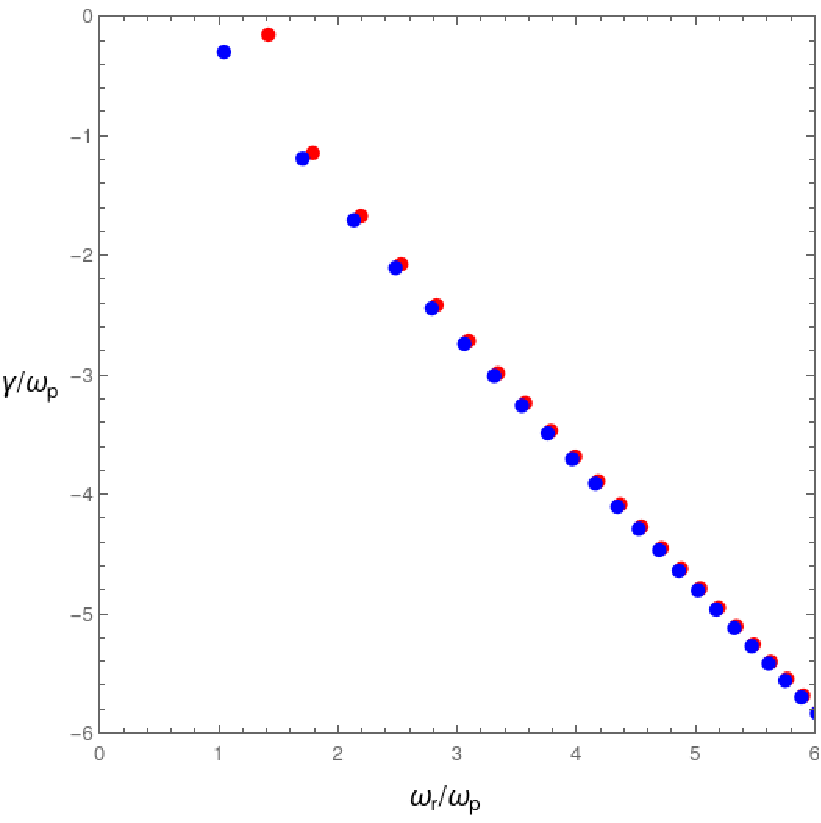}
\caption{Distributions of the exact values (red dots) and the approximate values (\diffColor dots) of the complex eigenfrequencies 
 for $k\lambda_D = 1/2$ in the region where $\omega_r  > 0$. 
 Here,  Eqs.~(\ref{zeta_n}) and (\ref{zeta2}) with $n=0, 1, 2, \cdots$ are 
used to evaluate the approximate eigenfrequencies. 
}
\label{fig:omegaAlpha2}
\end{figure}

In Fig.~\ref{fig:omegaAlpha2}, 
the exact complex eigenfrequencies and  
the approximate eigenfrequencies 
 for $k\lambda_D = 1/2$ are plotted as red and \diffColor dots, 
respectively, in the region where ${\rm Re}(\omega) > 0$. 
Here,  Eqs.~(\ref{zeta_n}) and (\ref{zeta2}) with $n=0, 1, 2, \cdots$ are 
used to evaluate the approximate eigenfrequencies. 
It is observed that the approximate values approach the exact values 
as the magnitude of the complex eigenfrequencies increases.

\subsection{
Approximate expression for effects of an infinite number of complex-valued eigenfrequencies
}

We here consider an infinite series, 
$
\sum_{n= N}^\infty
{\rm Re}
[ e^{i \zeta_n \tau} / (C - \zeta_n^2) ]
$,
where $C$ is a positive constant and $\tau \equiv k v_T t$.  
This form of the infinite series is included in Eq.~(\ref{solE}) for $E(k, t)$. 
When $n \gg 1$ and $|\zeta_n| \gg C$, 
we use Eqs.~(\ref{zeta_n}) and (\ref{zeta2}) to 
 make the following approximations, 
\begin{eqnarray}
\label{ReeC}
&&{\rm Re}
\left[
\frac{e^{i \zeta_n \tau}}{C - \zeta_n^2}
\right]\nonumber\\
&\simeq&  
- \frac{1}{|\zeta_n| ^2} \exp (|\zeta_n| \tau \sin \theta_n)
\cos (2 \theta_n + |\zeta_n | \tau \cos \theta_n )\nonumber
\\
&\simeq& 
- \frac{1}{|\zeta_n| ^2}  \exp \left( - \frac{|\zeta_n| \tau}{\sqrt{2}}\right)
\sin \left(\delta_n +\frac{|\zeta_n| \tau}{\sqrt{2}} \right)
,
\end{eqnarray}
where 
\begin{equation}
|\zeta_n |
\equiv 
\sqrt{ 
\left(
 2  n + 
 \frac{3}{4} 
\right) \pi
},
\hspace*{5mm}
\delta_n 
\equiv
2\theta_n + \frac{\pi}{2} 
\equiv
\frac{1}{ |\zeta_n|^2}
\log \left(
 \frac{\sqrt{\pi}}{\kappa^2}| \zeta_n |
\right)
.
\end{equation}
For $\tau = 0$,  using Eq.~(\ref{ReeC}) and $|\delta_n| \ll 1$ 
gives 
\begin{equation}
{\rm Re}
\left[
\frac{1}{C - \zeta_n^2}
\right]
\simeq  
- \frac{\delta_n}{|\zeta_n| ^2} 
.
\end{equation}
For $\tau > 0$, we can take $N \gg 1$ such that 
$|\zeta_{N}| \tau / \sqrt{2} \gg \delta_{N}>0$. 
Then, Eq.~(\ref{ReeC}) is rewritten as 
\begin{equation}
{\rm Re}
\left[
\frac{e^{i \zeta_n \tau}}{C - \zeta_n^2}
\right]
\simeq 
- \frac{1}{|\zeta_n| ^2}  \exp \left( - \frac{|\zeta_n| \tau}{\sqrt{2}}\right)
\sin \left(\frac{|\zeta_n| \tau}{\sqrt{2}} \right)
\end{equation}
where  $n \geq N$ and  $|\zeta_n| ^2 \propto n$. 
We find that, for a large but fixed integer $N$,  
\begin{eqnarray}
S_N^{N'} (\tau) 
&\equiv&
\sum_{n=N}^{N'}
{\rm Re}
\left[
\frac{e^{i \zeta_n \tau}}{C - \zeta_n^2}
\right]\nonumber\\&
\simeq&
- 
\sum_{n=N}^{N'}
\frac{1}{|\zeta_n| ^2}  \exp \left( - \frac{|\zeta_n| \tau}{\sqrt{2}}\right)
\sin \left(\frac{|\zeta_n| \tau}{\sqrt{2}} \right)
\end{eqnarray}
does not uniformly converge to 
$S_N (\tau) \equiv \lim_{N'\rightarrow \infty}S_N(\tau)$ because 
$| S_N^{N'} (\tau) - S_N (\tau) | < \epsilon$ 
requires $N' \geq N'(\epsilon, \tau)$ 
where 
$N'(\epsilon, \tau) \propto |\zeta_{N'(\epsilon, \tau)}|^2  
\propto \delta (\epsilon) / \tau^2$ 
which diverges as $\tau \rightarrow +0$. 
Therefore, the number of the complex-valued eigenfrequencies, which are 
necessary to be included for accurately evaluating $E(k, t)$,  diverges to infinity 
as $t \rightarrow +0$. 
On the other hand, 
when $v$ is fixed, the series expansion for representing $f_1(k, v, t)$ uniformly converges. 
It is because the extra factor $1/(\omega_n - k v)$ included in the series expansion 
accelerates the convergence as $n \rightarrow \infty$. 
Now, we use 
\begin{eqnarray}
2 \pi 
&=&
|\zeta_{n+1}|^2 - |\zeta_n|^2
= 
(|\zeta_{n+1}| + |\zeta_n|) (|\zeta_{n+1}| - |\zeta_n|) \nonumber\\
&\simeq &
2 
 |\zeta_n| (|\zeta_{n+1}| - |\zeta_n|) 
\end{eqnarray}
to write 
\begin{eqnarray}
\label{ReeC2}
&&
\hspace*{-8mm}
\sum_{n= N}^\infty
{\rm Re}
\left[
\frac{e^{i \zeta_n \tau}}{C - \zeta_n^2}
\right]
 \simeq  
- \sum_{n= N}^\infty
\frac{1}{|\zeta_n| ^2}  \exp \left( - \frac{|\zeta_n| \tau}{\sqrt{2}}\right)
\sin \left(\frac{|\zeta_n| \tau}{\sqrt{2}} \right)\nonumber
\\ 
& \simeq & 
-  \frac{1}{\pi}
\sum_{n= N}^\infty
\frac{(|\zeta_{n+1}| - |\zeta_n|) }{|\zeta_n|} 
\exp \left( - \frac{|\zeta_n| \tau}{\sqrt{2}}\right)
\sin \left(\frac{|\zeta_n| \tau}{\sqrt{2}} \right)
\nonumber
\\ 
& \simeq & 
-  \frac{1}{\pi}
\int_{|\zeta_{N}|}^{+\infty}
 \frac{d |\zeta |}{|\zeta |}
\exp \left( - \frac{|\zeta| \tau}{\sqrt{2}}\right)
\sin \left(\frac{|\zeta| \tau}{\sqrt{2}} \right)
 \nonumber
\\ 
& = & 
-  \frac{1}{\pi}
\int_{|\zeta_{N}|\tau/\sqrt{2}}^{+\infty}
dx \frac{e^{-x}\sin x}{x}\nonumber\\
&=& 
- \frac{1}{4}
+
\frac{1}{\pi}
\int_{0}^{|\zeta_{N}|\tau/\sqrt{2}}
dx \frac{e^{-x}\sin x}{x}
,
\end{eqnarray}
where 
\begin{equation}
\int_{0}^{+\infty}
dx \frac{e^{-x}\sin x}{x}
= \frac{\pi}{4}
\end{equation}
 is used. 

Now, using Eqs.~(\ref{solE}), (\ref{ReeC2}), and 
\begin{equation}
{\rm Re} \biggl[
\frac{e^{-i \omega_\mu t}
}{1 +  \kappa^2 - \omega_\mu^2/\omega_p^2}
\biggr]
= 
\frac{1}{2 \kappa^2} 
{\rm Re} \biggl[
\frac{e^{-i \zeta_\mu \tau}
}{(1 +  \kappa^2)/2\kappa^2- \zeta_\mu^2}
\biggr]
,
\end{equation}
we can express $E(k, t)$ as 
\begin{eqnarray}
&&E(k, t)
 \nonumber \\ & = &
-  i \; \alpha
\frac{m \omega_p^2}{e k} 
 ( 1+ \kappa^2 )
\sum_{{\rm Re} \zeta_\mu >0}
{\rm Re} \biggl[
\frac{e^{-i \zeta_\mu \tau}
}{(1 +  \kappa^2)/2\kappa^2- \zeta_\mu^2}
\biggr]\nonumber
\\ & \simeq &
-  i \; \alpha
\frac{m \omega_p^2}{e k}  ( 1+ \kappa^2 )\nonumber
\\ & & \mbox{} 
\times \biggl(
\sum_{0 < {\rm Re} \zeta_\mu < {\rm Re}\zeta_{N}}
{\rm Re} \biggl[
\frac{e^{-i \zeta_\mu \tau}
}{(1 +  \kappa^2)/2\kappa^2- \zeta_\mu^2}
\biggr]
\nonumber\\&&-  \frac{1}{\pi}
\int_{|\zeta_{N}| \tau/\sqrt{2}}^{+\infty}
dx \frac{e^{-x}\sin x}{x}
\biggr)\nonumber
\\
& \equiv &
E_{N} (k, t)
,
\label{E_oa_infty}
\end{eqnarray}
where $\zeta_n \equiv \omega_n / k v_T$ and 
$\zeta_n \tau \equiv \omega_n t$. 
Recall that we need to choose $N \gg 1$ such that 
$|\omega_{N}| t / \sqrt{2} \gg \delta_{N}>0$. 
However, 
when $N \gg 1$, 
$E (k, t=0) \simeq E_{N} (k, t=0) =
 i (4\pi n_0 e /k) \alpha  
= i (m/e) (\omega_p/k) \alpha$ 
also holds (as shown later) 
and $E (k, t) \simeq E_{N} (k, t)$ for any $t \geq 0$. 

From the residue theorem, we obtain
\begin{eqnarray}
\label{residue}
&&\frac{1}{2\pi}
\int_{-\infty}^{+\infty} d \zeta
\frac{Z(\zeta) e^{- i \zeta \tau}}{1 + \kappa^{-2}
[ 1 + \zeta Z(\zeta) ]}
\nonumber\\&=&
- i \sum_\mu 
\frac{ \kappa^2 ( 1+ \kappa^2) e^{-i \omega_\mu t}
}{1 +  \kappa^2 - \omega_\mu^2/\omega_p^2}
\nonumber\\&=& 
- 2 i \sum_{{\rm Re} \omega_\mu > 0}{\rm Re}
\biggl[
\frac{ \kappa^2 ( 1+ \kappa^2) e^{-i \omega_\mu t}
}{1 +  \kappa^2 - \omega_\mu^2/\omega_p^2}
\biggr]
\end{eqnarray}
where $\tau > 0$. 
Using the asymptotic expansion of $Z(\zeta)$ for $|\zeta| \gg 1$ and 
taking the limit for $\tau \rightarrow +0$, we find
\begin{eqnarray}
\label{lim0int}
&&\lim_{\tau \rightarrow + 0} 
\frac{1}{2\pi}
\int_{-\infty}^{+\infty} d \zeta
\frac{Z(\zeta) e^{- i \zeta \tau}}{1 + \kappa^{-2}
[ 1 + \zeta Z(\zeta) ]}\nonumber\\
&=&
- 2 i \lim_{\tau \rightarrow + 0} 
\sum_{{\rm Re} \omega_\mu > 0}{\rm Re}
\biggl[
\frac{ \kappa^2 ( 1+ \kappa^2) e^{-i \omega_\mu t}
}{1 +  \kappa^2 - \omega_\mu^2/\omega_p^2}
\biggr]
= i
.
\end{eqnarray}
However, for $\tau = 0$, we obtain 
\begin{eqnarray}
\label{tau0int}
& &
\frac{1}{2\pi}
\int_C d \zeta
\frac{Z(\zeta)}{1 + \kappa^{-2}
[ 1 + \zeta Z(\zeta) ]}
\nonumber \\ 
&=&
- 2 i 
\sum_{{\rm Re} \omega_\mu > 0}{\rm Re}
\biggl[
\frac{ \kappa^2 ( 1+ \kappa^2) 
}{1 +  \kappa^2 - \omega_\mu^2/\omega_p^2}
\biggr]
=  \frac{i}{4}  (3 - \kappa^2)
,
\end{eqnarray}
\textcolor{\diffColor}{
which is derived using the residue theorem. 
Appendix~B presents supplementary explanations about contours used for the integrals in Eqs.(66), (67), and (68).
In Eq.~(\ref{tau0int}), the integral $\int_C d \zeta$ is done 
along the circle $C: \zeta = R e^{i \theta}$ in the complex $\zeta$-plane 
where the radius of $C$ is given by $R\gg 1$ and the orientation of $C$ is taken clockwise by 
varying the argument $\theta$ of $\zeta$
from $\pi$ to $-\pi$. }
Comparing Eqs.~(\ref{lim0int}) and (\ref{tau0int}) shows that changing the order of 
operations $\lim_{\tau \rightarrow + 0}$ and $\sum_\mu$ results in 
different values. 
Thus, when any large but finite fixed number of eigenfrequencies 
are included to evaluate 
$E(k, t)$, the limit value for  $\tau \rightarrow + 0$ does not 
converge to the correct initial value but to another value, 
the ratio of which to the correct one is estimated from 
Eqs.~(\ref{lim0int}) and (\ref{tau0int}) as 
$\frac{1}{4}  (3 - \kappa^2)$. 
We use Eq.~(\ref{tau0int}) to confirm that 
the approximate solution $E_{N}(k, t)$ given in Eq.~(\ref{E_oa_infty}) 
can correctly approach to the correct initial value  as $t \rightarrow +0$, 
\begin{eqnarray}
\lim_{N \rightarrow \infty}
E_{N} (k, t=0)
& =  & 
-  i \; \alpha
\frac{m \omega_p^2}{e k}  ( 1+ \kappa^2 )\nonumber
\\ & & \mbox{} 
\times \biggl(
\sum_{{\rm Re} \zeta_\mu > 0}
{\rm Re} \biggl[
\frac{1}{(1 +  \kappa^2)/2\kappa^2- \zeta_\mu^2}
\biggr]\nonumber\\
&&-  \frac{1}{\pi}
\int_0^{+\infty}
dx \frac{e^{-x}\sin x}{x}
\biggr)\nonumber
\\ & =  & 
i \alpha
\frac{m \omega_p^2}{e k} 
\biggl[  \frac{1}{4}  (3 - \kappa^2)
+ \frac{1}{4} ( 1+ \kappa^2 )
\biggr]\nonumber
\\
& = &
i  \alpha \frac{m \omega_p^2}{e k}
=
 i \alpha   \frac{4\pi n_0 e}{k} \nonumber\\ &=& 
E (k, t=0) 
\end{eqnarray}

\begin{figure}[h]
\includegraphics[width=8cm]{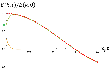}
\includegraphics[width=8cm]{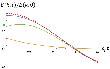}
\caption{
Time evolutions of the normalized electric field $E(k, t)/E(k, 0)$ for $k\lambda_D=1/2$.
The top panel shows 
$E_{<  N}(k,t)$ (green line), $E_{> N}(k,t)$  (orange line), 
and $E_{N}(k,t) = E_{<  N}(k,t) + E_{> N}(k,t)$ (\diffColor line) 
for $N = 100$. 
%
%
The bottom panel shows $E_{<  N}(k,t)$ (green line), $E_{> N}(k,t)$  (orange line), 
and $E_{N}(k,t) = E_{<  N}(k,t) + E_{> N}(k,t)$ (\diffColor line) 
for $N = 1$. 
The red dotted lines in both panels represent the CD simulation results for $\alpha=0.01$.
The diamonds plotted in the vertical axes of the top and bottom panels 
correspond to the value given by 
$\lim_{N\rightarrow \infty} E_{<  N}(k,0) / E(k, 0) = ( 3 - \kappa^2 ) / 4 = 11/16 = 0.6875$ for $\kappa \equiv k \lambda_D = 1/2$. 
}
\label{fig:EtVer}
\end{figure}

Now, we express the approximate solution $E_{N} (k, t)$ in Eq.(\ref{E_oa_infty}) by  
the sum of two parts as 
\begin{equation}
E_{N} (k, t)
=
E_{<  N}(k,t)+E_{> N}(k,t)
,
\end{equation}
where 
\begin{eqnarray}
E_{<  N}(k,t)
& \equiv &
-  i \; \alpha
\frac{m \omega_p^2}{e k}  ( 1+ \kappa^2 )\nonumber
\\ & & \mbox{} 
\times 
\sum_{0 < {\rm Re} \zeta_\mu < {\rm Re}\zeta_{N}}
{\rm Re} \biggl[
\frac{e^{-i \zeta_\mu \tau}
}{(1 +  \kappa^2)/2\kappa^2- \zeta_\mu^2}
\biggr]
,
\nonumber \\
E_{> N}(k,t)
& \equiv &
  i \; \alpha
\frac{m \omega_p^2}{e k}  ( 1+ \kappa^2 )\nonumber
\\ & & \mbox{} 
\times   \frac{1}{\pi}
\int_{|\zeta_{N}| \tau/\sqrt{2}}^{+\infty}
dx \frac{e^{-x}\sin x}{x}
.
\label{EtVer}
\end{eqnarray}
The top panel of Fig.~\ref{fig:EtVer} shows $E_{<  N}(k,t)$, $E_{> N}(k,t)$, and $E_{N}(k,t) = E_{<  N}(k,t) + E_{> N}(k,t)$, 
for $N = 100$. 
There, the approximate solution $E_{N}(k,t)$ for $N=100$ and the CD simulation results 
for $\alpha = 0.01$ are plotted 
by \diffColor solid and red dashed lines, respectively. 
Since the effects of complex eigenfrequencies $\omega_\mu$ with large absolute values included in $E_{> N}(k,t)$ 
become more significant as $t$ approaches 0, 
$E_{<  N}(k,t)$ deviates from the CD simulation results for small $t$ (specifically $t < 0.2$ in Fig.~\ref{fig:EtVer}). 
The diamonds plotted in the vertical axes of the top and bottom panels in Fig.~\ref{fig:EtVer} 
correspond to the value given by 
$\lim_{N\rightarrow \infty} E_{<  N}(k,0) / E(k, 0) = ( 3 - \kappa^2 ) / 4 = 11/16 = 0.6875$ for $\kappa \equiv k \lambda_D = 1/2$. 
However, the approximate solution agrees with the CD simulation result by adding $E_{> N}(k,t)$. 
The bottom panel shows $E_{<  N}(k,t)$, $E_{> N}(k,t)$, and $E_{N}(k,t) = E_{<  N}(k,t) + E_{> N}(k,t)$ for $N =1$. 
Even in this case of $N =1$, 
the difference between $E_{N}(k,t)$ and the CD simulation result near $t=0$ is as small as about 5\%.

As described in Appendix~D, 
the Vlasov-Poisson system is symmetric under the time reversal transformation 
whether the linear approximation is made or not. 
Especially, under the initial condition made in this section, 
the distribution function at $t=0$ is an even function of $v$, 
from which we find that the distribution function is an even function of 
time $t$ as explained in Appendix~D. 
Accordingly, 
$f_1 (k, -v, -t) = f_1( k ,v, t )$ and $E(k, -t) = E(k, t)$ hold 
for any $t$ and $v$. 
Therefore, 
$E(k, t)$ and $f_1( k ,v, t )$ for $t < 0$ 
are immediately given from Eqs.~(\ref{solE}) and (\ref{solf1}) with 
making use of the time reversal.

\subsection{Approximate solution in early time}

We here use the asymptotic expansion of $Z(\zeta)$  for $|\zeta| \gg 1$ given in 
Eq.~(\ref{asymptoticZ}) to derive 
\begin{eqnarray}&&
1 + \kappa^{-2} [ 1 + \zeta Z(\zeta) ]\nonumber\\
&=&
1 - \kappa^{-2}
\sum_{n=1}^{N-1}
\frac{\Gamma (n + 1/2) }{\sqrt{\pi} } \frac{1}{\zeta^{2n}}\nonumber\\
&+& \kappa^2 \frac{\Gamma (N + 1/2) }{\sqrt{\pi} } \zeta e^{-\zeta^2} 
\int_{i\infty}^\zeta \frac{e^{s^2}}{s^{2N}} ds 
.
\end{eqnarray}
Then, for $|\zeta| \gg 1$ and 
$- \pi / 4 < \arg \zeta < 5 \pi / 4$, 
we obtain
\begin{eqnarray}
\label{asymptotic2}
&&
\frac{Z(\zeta)}{1 + \kappa^{-2} [ 1 + \zeta Z(\zeta) ]}\nonumber\\
&=&
- \frac{1}{\zeta}
\left[
1 + 
\sum_{n=1}^{N} e_n (\kappa^2) \zeta^{-2n}
+ {\cal O}(\zeta^{-2N-2})
\right]
,
\end{eqnarray}
where  $e_n (\kappa^2)$ $(n = 1, 2, \cdots)$ are recursively defined by 
\begin{equation}
e_n (\kappa^2) \equiv 
\frac{\Gamma(n+1/2)}{\Gamma(1/2)}
+ \kappa^{-2} 
\sum_{j = 0}^{n-1} \frac{\Gamma(n-j+1/2)}{\Gamma(1/2)} e_j (\kappa^2) 
,
\end{equation}
with $e_0 (\kappa^2) \equiv 1$. 
Thus, we find 
\begin{eqnarray}
e_1(\kappa^2) 
&  =  & \frac{1}{2}  + \frac{1}{2} \kappa^{-2}
,
\nonumber \\ 
e_2(\kappa^2) 
& \equiv & 
\frac{3}{4}  + \kappa^{-2}  \left( \frac{3}{4}  + \frac{1}{2}   e_1(\kappa^2)  \right)
= 
\frac{3}{4}  + \kappa^{-2} +  \frac{1}{4} \kappa^{-4}, 
\nonumber \\ 
\cdots .
&  
&
\end{eqnarray}
We now use Eq.~(\ref{asymptotic2}) to get 
\begin{eqnarray}
\label{asymptotic3}
&&\int_C \frac{d \zeta}{2\pi}
\frac{Z(\zeta) e^{- i \zeta \tau}}{1 + \kappa^{-2}
[ 1 + \zeta Z(\zeta) ]}\nonumber\\
& = & 
-  
\int_C \frac{d \zeta}{2\pi \zeta}
\biggl[
\sum_{n=0}^N
\frac{e_n(\kappa^2)}{\zeta^{2n}}
+ {\cal O}(\zeta^{-2N-2} )
\biggr] e^{-i \zeta \tau} \nonumber
\\ 
& = & 
i \sum_{n=0}^N \frac{(-1)^n}{(2n)!} e_n(\kappa^2) \tau^{2n}
+ {\cal O}(\tau^{2N+2} )
,
\end{eqnarray}
where the integral $\int_C d \zeta$ is done 
along the circle $C: \zeta = R e^{i \theta}$ in the complex $\zeta$-plane. 
Here, the radius of $C$ is given by $R\gg 1$ and the orientation of $C$ is taken clockwise by 
varying the argument $\theta$ of $\zeta$ 
from $\pi$ to $-\pi$. 
\textcolor{\diffColor}{
Supplementary explanations about the derivation of Eq.~(\ref{asymptotic3}) are given after Eq.~(\ref{Aasymptotic}) in Appendix~B.
}
Equation~(\ref{solE}) is rewritten here for $\tau \equiv k v_T t > 0$ as 
\begin{eqnarray}
\label{Ekt2}
E(k, t)
&=&
\alpha
\frac{4 \pi e n_0}{k}
\int_C \frac{d \zeta}{2\pi} \;
\frac{Z(\zeta)e^{-i \zeta \tau}}{1 + \kappa^{-2}
[ 1 + \zeta Z(\zeta) ]}
, 
\end{eqnarray}
into which Eq.~(\ref{asymptotic3}) is substituted to derive 
the Taylor expansion of $E(k, t)$ about $\tau = k v_T t = 0$ is written as 
\begin{equation}
E(k, t) 
=
i \alpha
\frac{4 \pi e n_0}{k}
\sum_{n = 0}^N
 \frac{(-1)^n }{(2n)!}
e_n (\kappa^2) \tau^{2n}
+ {\cal O}(\tau^{2N+2} )
.
\label{EtTau}
\end{equation}

Next, Eq.~(\ref{solf1}) for $f_1(k,v,t)$ is rewritten here as
\begin{eqnarray}
\label{f1kvt2}
f_1(k, v, t)
=&&
\alpha f_0 (v) 
\Biggl[ 
\exp \left( - i \frac{v}{v_T} \tau \right)
- i \kappa^{-2}\frac{v}{v_T}
\int_{-\infty}^{+\infty}\nonumber\\&& \frac{d\zeta}{2\pi}  
\; 
\frac{Z(\zeta)e^{-i \zeta \tau}}{(\zeta - v / v_T ) \{ 1 + \kappa^{-2}
[ 1 + \zeta Z(\zeta) ]\} }
\Biggr]
. 
\end{eqnarray}
Then, we use 
\begin{equation}
\frac{1}{\zeta - v / v_T }
= 
\frac{1}{\zeta}
 \left[
1 + \sum_{n=1}^\infty 
\left( \frac{v}{v_T} \right)^n \zeta^{-n} 
\right]
,
\end{equation}
and 
Eq.~(\ref{asymptotic3}) to obtain 
\begin{eqnarray}
\label{asymptotic4}
&&\frac{Z(\zeta)}{(\zeta - v / v_T ) \{ 1 + \kappa^{-2}
[ 1 + \zeta Z(\zeta) ]\} }\nonumber\\
&=&
- \frac{1}{\zeta^2}
 \left[
1 + \sum_{n=1}^{N}
d_n (\kappa^2, v/v_T) \zeta^{-n}
+ {\cal O}(\zeta^{-N-1} )
\right]
, 
\end{eqnarray}
where $d_n (\kappa^2, v/v_T)$ $(n=0, 1, 2, \cdots)$ are defined by 
\begin{equation}
\label{dn}
d_n (\kappa^2, v/v_T) 
\equiv
\sum_{j=0}^{[n/2]} (v/v_T)^{n-2j} e_j (\kappa^2)
, 
\end{equation}
and $[n/2]$ denotes the greatest integer less than or equal to $n/2$. 
From Eq.~(\ref{dn}), we have 
\begin{eqnarray}
d_0 (\kappa^2, v/v_T) 
&  =  & 1
\nonumber \\ 
d_1 (\kappa^2, v/v_T) 
&  =  & v/v_T
\nonumber \\ 
d_2 (\kappa^2, v/v_T) 
& = & 
( v/v_T)^2 + e_1 (\kappa^2)
\nonumber \\ 
d_3 (\kappa^2, v/v_T) 
& = & 
( v/v_T)^3 + ( v/v_T) \, e_1 (\kappa^2)
\nonumber \\ 
d_4 (\kappa^2, v/v_T) 
& = & 
( v/v_T)^4 + ( v/v_T)^2 \, e_1 (\kappa^2) + e_2 (\kappa^2)
\nonumber \\ 
d_5 (\kappa^2, v/v_T) 
& = & 
( v/v_T)^5 + ( v/v_T)^3 \, e_1 (\kappa^2) + ( v/v_T) \, e_2 (\kappa^2)
\nonumber \\ 
\cdots
.
&  & 
\end{eqnarray}
Integrating Eq.~(\ref{asymptotic4}) with respect to $\zeta$ yields 
\begin{eqnarray}
\label{asymptotic5}
&&
\int_{-\infty}^{+\infty} \frac{d\zeta}{2\pi}  
\; 
\frac{Z(\zeta)e^{-i \zeta \tau}}{(\zeta - v / v_T ) \{ 1 + \kappa^{-2}
[ 1 + \zeta Z(\zeta) ]\} }
\nonumber\\
 & = &
-
\int_C  \frac{d\zeta}{2\pi}  
\; 
\frac{e^{-i \zeta \tau}}{\zeta^2}
 \biggl[
1 + \sum_{n=1}^{N}
d_n (\kappa^2, v/v_T) \zeta^{-n}
+ {\cal O}(\zeta^{-N-1} )
\biggr]\nonumber
\\ & = &
i \sum_{n=0}^{N}
\frac{(-i \tau)^{n+1}}{(n+1)!}
d_n (\kappa^2, v/v_T) 
+ {\cal O}(\tau^{N+2} )
.
\end{eqnarray}
Combining Eqs.~(\ref{f1kvt2}) and (\ref{asymptotic5}), 
the Taylor expansion of $f_1(k, v, t)$ about $\tau = k v_T t = 0$ 
is derived as 
\begin{eqnarray}
f_1(k, v, t)
& = & 
\alpha f_0 (v) 
\biggl[ 
\exp \left( - i \frac{v}{v_T} \tau \right)\nonumber\\
&&+ \kappa^{-2}\frac{v}{v_T}
\sum_{n=1}^{N}
\frac{(-i \tau)^n}{n!}
d_{n-1} (\kappa^2, v/v_T) 
+ {\cal O}(\tau^{N+1} )
\biggr]\nonumber
\\ 
& = & 
\alpha f_0 (v) 
\biggl[ 
1+ 
\sum_{n=1}^{N}
\frac{(-i \tau)^n}{n!}
\biggl\{
\left( \frac{v}{v_T} \right)^n\nonumber\\
&&+ \kappa^{-2}\frac{v}{v_T}
d_{n-1} (\kappa^2, v/v_T) 
\biggr\}
+ {\cal O}(\tau^{N+1} )
\biggr]\nonumber
\\ 
& = & 
\alpha f_0 (v) 
\biggl[ 
1+ 
\sum_{n=1}^{N}
\frac{(-i \tau)^n}{n!}
f_n (\kappa^2, v/v_T) \nonumber\\
&&+ {\cal O}(\tau^{N+1} )
\biggr]
,
\label{f1Tau}
\end{eqnarray}
where $f_n (\kappa^2, v/v_T)$ $(n=1, 2, \cdots)$ 
are defined by
\begin{equation}
f_n (\kappa^2, v/v_T) 
\equiv 
\left( \frac{v}{v_T} \right)^n
+ \kappa^{-2}\frac{v}{v_T}
d_{n-1} (\kappa^2, v/v_T) 
.
\end{equation}

We now compare 
the analytical solutions obtained by the series expansions 
in Eqs.~(\ref{EtTau}) and (\ref{f1Tau}) with the CD simulation results 
for $\alpha =0.01$.
\begin{figure}[h]
\centering
\includegraphics[width=8cm]{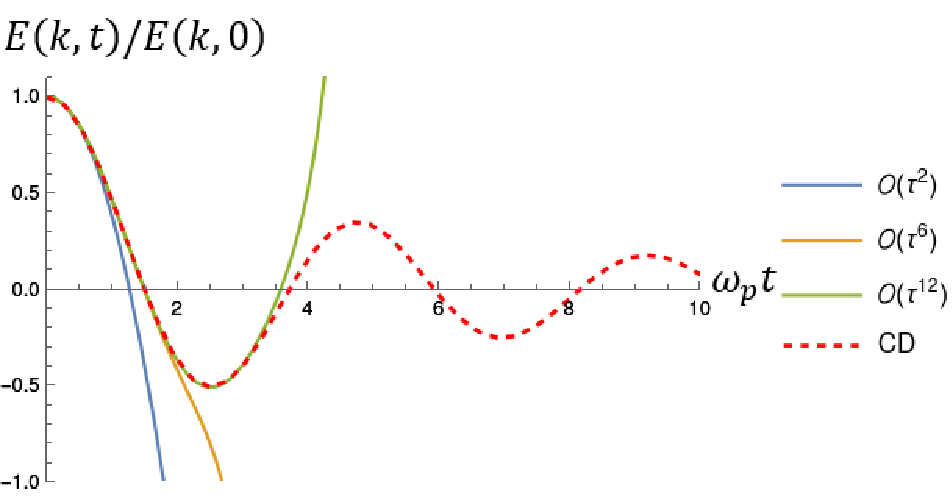}
\caption{Time evolution of $E(k, t)/E(k, 0)$ for $k\lambda_D = 1/2$
The red dashed line represents the CD simulation result for $\alpha =0.01$.
Results obtained from Eq.~(\ref{EtTau}) including terms up to the orders of $\tau^2$, 
$\tau^6$, and $\tau^{12}$ are shown by the \diffColor, orange, and green lines, respectively. 
}
\label{fig:ETau}
\end{figure}
Figure~\ref{fig:ETau} shows the analytical solutions of Eq.~(\ref{EtTau}) for $k \lambda_D = 1/2$ 
including terms up to orders of $\tau^2, \tau^6$, and $\tau^12$. 
We can confirm that the discrepancy from the simulation results decreases 
as the number of the included terms increases. 
The calculation up to $\tau^6$ fits well with the CD simulation results when 
$\omega_p t < \sqrt{2}$ (which corresponds to 
$\tau \equiv k v_T t \equiv \sqrt{2} (k \lambda_D) \omega_p t <1$).
Figure~\ref{fig:f1Tau13} shows the plot of $f_1(x,v,t)/f_0(v)$ for $k \lambda_D = 1/2$ calculated using Eq.~(\ref{f1Tau}) 
with terms up to the order of $\tau^{13}$ included. 
It shows a good agreement with the CD simulation results shown in the bottom row of Fig.~\ref{fig:f1} 
for $\omega_p t \leq1$.

\begin{figure*}[ht]
\centering
\includegraphics[width=18cm]{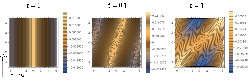}
\caption{Contours of $f_1(x,v,t)/( \alpha^2 f_0(v) )$ in the $(x, v)$-space for $k \lambda_D = 1/2$ calculated 
using Eq.~(\ref{f1Tau}) 
with terms up to the order of $\tau^{13}$ included. 
It shows a good agreement with the CD simulation results shown in the bottom row of Fig.~\ref{fig:f1} for 
$\omega_p t \leq1$.
}
\label{fig:f1Tau13}
\end{figure*}
%
\twocolumngrid

\section{
Analytical solution based on quasilinear theory and its verification by numerical simulation
}

\subsection{Spatially averaged distribution function}

In this subsection, formulas that hold rigorously 
for the spatially averaged distribution function and its associated entropy are 
derived without using linear approximations based on small perturbation amplitudes. 
The distribution function $f(x, v, t)$ is a periodic function of $x$ and 
the period length is given by $L \equiv 2\pi/k$. 
We use 
\begin{equation}
\langle \cdots \rangle \equiv 
\frac{1}{L} \int_{-L/2}^{L/2} \cdots \; dx 
,
\end{equation}
to denote the average with respect to $x$. 
The Vlasov equation is averaged in $x$ to yield
\begin{equation}
\label{ave-Vlasov}
\frac{\partial}{\partial t}
 \langle f \rangle  (v, t)
+
\frac{\partial}{\partial v}
\left[
\langle f \rangle  (v, t) \, a (v, t) 
\right]
=
0
,
\end{equation}
where $\langle f \rangle  (v, t) \equiv \langle f (x, v, t) \rangle $ is 
the electron distribution function averaged in $x$, 
and 
$a (v, t)$ represents the average acceleration of electrons defined by 
\begin{equation}
\langle f \rangle  (v, t) \,   a (v, t) 
\equiv 
- \frac{e}{m} \langle E(x, t) \,  f(x, v, t) \rangle
.
\end{equation}
From Eq.~(\ref{ave-Vlasov}), we have 
\begin{eqnarray}
& & 
\frac{\partial}{\partial t}
\left[ \langle f \rangle  (v, t) \frac{1}{2} m v^2 \right]
+
\frac{\partial}{\partial v}
\left[
\langle f \rangle  (v, t)  \frac{1}{2} m v^2 a (v, t) 
\right]
\nonumber \\ & & 
=
\langle f \rangle  (v, t)  \, m \, v \,
a (v, t) 
,
\end{eqnarray}
and 
\begin{equation}
\frac{d}{dt} 
\int_{-\infty}^{+\infty} dv \; 
\langle f \rangle  (v, t) \frac{1}{2} m v^2
= 
\int_{-\infty}^{+\infty} dv \; 
\langle f \rangle  (v, t)  \, m \, v \,
a (v, t) 
,
\end{equation}
where $m \, v \, a (v, t)$ represents the $x$-averaged rate of change of the kinetic energy of 
an electron with velocity $v$ at time $t$ caused by the electric field. 
The total energy conservation of the Vlasov-Poisson system is written as 
\begin{equation}
\frac{d}{dt} 
\left[
\int_{-\infty}^{+\infty} dv \; 
\langle f \rangle  (v, t) \frac{1}{2} m v^2
+ \frac{1}{8\pi}  \langle E^2  \rangle
\right]
= 0
.
\end{equation}
The Gibbs entropy $S[f]$ per unit length in the $x$-direction is defined as a functional of the 
distribution function $f(x, v, t)$ by 
\begin{equation}
S[f]
\equiv 
- \left\langle 
\int_{-\infty}^{+\infty} dv \; 
f(x, v, t) \log f(x, v, t)
\right\rangle
,
\end{equation}
which is found to be an invariant,
\begin{equation}
\frac{d}{dt} S[f]
= 0
.
\end{equation}
We now use the $x$-averaged distribution function $\langle f \rangle (v, t)$ to 
define the entropy density $S[\langle f \rangle]$ by 
\begin{equation}
S[\langle f \rangle]
\equiv 
- 
\int_{-\infty}^{+\infty} dv \; 
\langle f \rangle (v, t) \log \langle f \rangle (v, t)
,
\end{equation}
which is not an invariant but a function of time $t$. 
From the viewpoint of information theory,~\cite{Information}
the entropy density $S[\langle f \rangle]$ is given from the average of 
\begin{equation}
S(v, t) 
\equiv 
-  \log \langle f \rangle (v, t)
= 
-  \log [\langle f \rangle (v, t) ( dv / n_0)]
+  \log (dv/n_0)
,
\end{equation}
which represents the Shannon information content (or self-entropy)
$-  \log [\langle f \rangle (v, t) ( dv / n_0)] = - \log [ P_V(v_j)] $ 
[see Eq.~(\ref{SEXV})] 
plus an additional constant given by $\log (dv/n_0)$ 
for the probability $\langle f \rangle (v, t) ( dv / n_0)$ of finding the electron velocity 
in the interval 
$[v - \frac{1}{2}dv, v + \frac{1}{2}dv]$ 
where $dv$ is regarded as an infinitesimal positive constant. 
Supplementary explanations on information entropies 
in the Vlasov-Poisson system are presented in Appendix~E. 
We now consider $S(v, t)$ as the information entropy (except an additional constant) 
of the electron with the velocity $v$ at time $t$. 
From Eq.~(\ref{ave-Vlasov}), we obtain 
\begin{eqnarray}
\label{dtS}
\left( \frac{\partial}{\partial t} + a(v, t) \frac{\partial}{\partial v} \right)
S(v, t)  
& = & 
- \left( \frac{\partial}{\partial t} + a(v, t) \frac{\partial}{\partial v} \right)
\log f(v, t)  
\nonumber \\ 
& = & 
 \frac{\partial a(v, t)}{\partial v}
.
\end{eqnarray}

Here, we define $u(v_0, t)$ which satisfies the differential equation, 
\begin{equation}
\frac{\partial}{\partial t} u(v_0, t)
=
a ( u(v_0, t), t) 
\end{equation}
with the initial condition 
\begin{equation}
u(v_0, t=0)
=
v_0
.
\end{equation}
Then, $u(v_0, t)$ represents the velocity of the electron which has 
the initial velocity $v_0$ and the history of acceleration 
$a(v, t')$ $(0\leq t' \leq t)$. 
The interval $[v_0 - \frac{1}{2}dv_0, v_0 + \frac{1}{2}dv_0]$ 
in the $v$-space evolves to the interval 
$[u (v_0, t) - \frac{1}{2}du (v_0, t) , u (v_0, t) + \frac{1}{2}du (v_0, t)]$ 
at time $t$  
when the velocities of the electrons in the interval 
are given by $u(v', t')$ 
($v_0 - \frac{1}{2}dv_0 \leq v' \leq v_0 + \frac{1}{2}dv_0$, 
 $0\leq t' \leq t)$. 
We note that the number (or probability) of the electrons found in the interval 
$[u (v_0, t) - \frac{1}{2}du (v_0, t) , u (v_0, t) + \frac{1}{2}du (v_0, t)]$ 
is invariant in time,
\begin{equation}
\langle f \rangle ( u(v_0, t),  t ), t )
\; du (v_0, t)
=
\langle f \rangle (v_0, 0)
\; dv_0
.
\end{equation}

Using Eqs.~(\ref{ave-Vlasov}) and (\ref{dtS}), we find 
\begin{eqnarray}
\frac{\partial }{\partial t} S(u(v_0, t), t)  
& =  & 
- \biggl[ \biggl( \frac{\partial }{\partial t}
 + a (v, t) \frac{\partial }{\partial v} 
\biggr) \log \langle f \rangle (v, t) 
\biggr]_{v= u(v_0, t)}
\nonumber \\ 
& =  & 
\biggl[
\frac{\partial a(v, t)}{\partial v}
\biggr]_{v= u(v_0, t)}
\end{eqnarray}
which implies that 
$\partial a(v, t)/\partial v$ represents the rate of change in 
the information entropy $S(v, t)  \equiv -  \log \langle f \rangle (v, t)$ 
of the electron with the velocity $v$ at time $t$ 
along the trajectory $u(v_0, t)$ in the $v$-space. 
Then, the increase of $S(v, t)$ along the trajectory during the 
time interval $[0, t]$ is given by 
\begin{eqnarray}
& & 
\Delta S(u(v_0, t), t)  
\equiv S (u(v_0, t), t)  - S (v_0, 0)
\nonumber \\ 
& & 
= - \log 
\biggl[ 
\frac{\langle f \rangle (u(v_0, t), t)}{\langle f \rangle (v_0, 0)}
\biggr]
=
\int_0^t dt'
\biggl[
\frac{\partial a(v, t')}{\partial v}
\biggr]_{v= u(v_0, t')}
,
\end{eqnarray}
from which we obtain 
\begin{eqnarray}
\label{fut}
& & 
\langle f \rangle (u(v_0, t), t) 
 =  
\langle f \rangle (v_0, 0) \; 
\exp 
\bigl[
- \Delta S(u(v_0, t), t)  
\bigr]
\nonumber \\ 
&  &
= 
\langle f \rangle (v_0, 0) \; 
\exp 
\biggl(
- \int_0^t dt'
\biggl[
\frac{\partial a(v, t')}{\partial v}
\biggr]_{v= u(v_0, t')}
\biggr)
.
\end{eqnarray}
We also find that the rate of change in the entropy of $S[\langle f \rangle]$ is given by 
\begin{eqnarray}
\frac{d}{d t} S[\langle f \rangle]  
& =  &
\frac{d}{d t} 
\int_{-\infty}^{+\infty} dv \; 
\langle f \rangle (v, t) \; 
S (v, t)
\nonumber  \\
& = &
\int_{-\infty}^{+\infty} dv \; 
\langle f \rangle (v, t) \;
\frac{\partial a(v, t)}{\partial v}
.
\end{eqnarray}

Another formula is obtained as 
\begin{equation}
 \frac{\partial }{\partial t} \log \langle f \rangle (u(v_0, t), 0) 
=
\biggl[ 
a (v, t) \frac{\partial }{\partial v} 
\log \langle f \rangle (v, 0) 
\biggr]_{v= u(v_0, t)}
,
\end{equation}
which is integrated in time $t$ to derive 
\begin{equation}
\log 
\biggl[ 
\frac{\langle f \rangle (u(v_0, t), 0)}{\langle f \rangle (v_0, 0)}
\biggr]
= 
\int_0^t dt'
\biggl[ 
a (v, t') \frac{\partial }{\partial v} 
\log \langle f \rangle (v, 0) 
\biggr]_{v= u(v_0, t')}
\end{equation}
and 
\begin{eqnarray}
\label{fu0}
& &  \langle f \rangle (u(v_0, t), 0)
\nonumber \\ 
&  & 
\hspace*{-3mm}
=
\langle f \rangle (v_0, 0)
 \exp \biggl(
\int_0^t dt'
\biggl[ 
a (v, t') \frac{\partial }{\partial v} 
\log \langle f \rangle (v, 0) 
\biggr]_{v= u(v_0, t')}
\biggr)
.
\hspace*{8mm}
\end{eqnarray}
Then, using Eqs.~(\ref{fut}) and (\ref{fu0}), we obtain 
\begin{equation}
\label{fut2}
 \log 
\biggl[ 
\frac{\langle f \rangle (u(v_0, t), t)}{\langle f \rangle (v_0, 0)}
\biggr]
=  
\Omega_t(v_0) 
,
\end{equation}
and 
\begin{equation}
\label{fut0}
\langle f \rangle (u(v_0, t), t) 
= \langle f \rangle (u(v_0, t),  0) \;
\exp \Omega_t (v_0)
,
\end{equation}
where the function $\Omega_t(v_0)$ is defined by 
\begin{eqnarray}
\label{Omegatv0}
\Omega_t(v_0) 
& \equiv & 
 \int_0^t dt' \; \Omega (u(v_0, t'), t')
, 
\end{eqnarray}
with
\begin{equation}
\label{Omegavt0}
\Omega (v, t)
\equiv 
- a (v, t) \frac{\partial }{\partial v} 
\log \langle f \rangle (v, 0) 
- \frac{\partial a(v, t)}{\partial v}
.
\end{equation}

Now, we consider the case in which the initial $x$-averaged distribution function 
$ \langle f \rangle (v, 0)  = f_0 (v)$
is given by the Maxwellian equilibrium distribution function in Eq.~(\ref{Maxwellian}). 
Then, we have 
\begin{equation}
\frac{\partial }{\partial v} 
\log f_0 (v) 
= - \frac{m \; v}{T}
,
\end{equation}
which is substituted into Eq.~(\ref{fu0}) to obtain 
\begin{equation}
f_0 (u(v_0, t))
= 
f_0(v_0)
\; \exp \biggl(
- \frac{\Delta {\cal E} (u(v_0, t), t)}{T}
\biggr)
,
\end{equation}
where
\begin{equation}
\Delta {\cal E} (u(v_0, t), t)
\equiv 
m
\int_0^t dt'
\; 
u(v_0, t') \; 
a (u(v_0, t'), t')
\end{equation}
represents the change in the kinetic energy of the electron with the initial velocity $v_0$ 
 caused by the acceleration due to the electric field during the time interval $[0, t]$. 
Then, we can rewrite Eqs.~(\ref{Omegavt0}), (\ref{Omegatv0}), and (\ref{fut0}) as
\begin{equation}
\label{Omegavt2}
\Omega (v, t)
=
\frac{m}{T}  v \; a (v, t) 
- \frac{\partial a(v, t)}{\partial v}
, 
\end{equation}
\begin{equation}
\label{Omegatv02}
\Omega_t  (v_0)
=
\frac{\Delta {\cal E} (u(v_0, t), t)}{T} - \Delta S (u(v_0, t), t)
, 
\end{equation}
and 
\begin{eqnarray}
\label{fexpOmega}
& & 
\langle f \rangle (u(v_0, t), t) 
=
 f_0 (u(v_0, t)) \;
\exp 
\Omega_t  (v_0)
\nonumber \\ & & 
= 
 f_0 (u(v_0, t)) 
\exp
\biggl[
\frac{\Delta {\cal E} (u(v_0, t), t)}{T} - \Delta S (u(v_0, t), t)
\biggr]
,
\hspace*{8mm}
\end{eqnarray}
respectively. 

Here, we should note that 
$\Omega_t  (v_0)$ defined in Eq.~(\ref{Omegatv02}) 
corresponds to the dissipation function employed 
by Evans and Searles to present their fluctuation theorem.~\cite{FT,Evans}
The fluctuation theorem by Evans and Searles is based on 
the time-reversible Liouville equation and gives 
the formula for the ratio of the probabilities that the dissipation 
function takes the positive and negative values with the same absolute value. 
The Vlasov equation differs from the Liouville equation treated by Evans and Searles 
in that it contains the electron's acceleration term determined from the distribution 
function through Poisson's equation.  
Therefore, the fluctuation theorem cannot be directly applied to 
the function $\Omega_t  (v_0)$ in our case. 
However, as explained in Appendix~E, 
we can show 
the non-negativity of the expected value of $\Omega_t  (v_0)$, 
which leads to the inequality in the form of the second law of thermodynamics. 

\subsection{Analysis to second order in perturbation amplitude}

In this subsection, we investigate 
the time evolution of the $x$-averaged 
distribution function considered in Sec.~IV.A 
up to the second order in the perturbation amplitude $\alpha$ 
based on the results of the linear analysis in Sec.~II. 
Recall that the initial condition for the distribution function is given by 
\begin{eqnarray}
\label{initial}
f(x, v, t=0)
& = & 
f_0 (v) + {\rm Re} [f_1 (k, v, t=0) \exp ( i \; k \;  x )]
\nonumber \\ 
& = &
f_0 (v) [ 1 + \alpha \cos (k \, x) ]
,
\end{eqnarray}
where $f_0(v)$  is the Maxwellian in Eq.~(\ref{Maxwellian}) 
and $f_1 (k, v, t=0) = \alpha f_0(v)$ is used. 
We have already derived the linear solution of the perturbed distribution function 
$f_1(x, v, t) = {\rm Re} [f_1 (k, v, t) \exp ( i \; k \;  x )]$ for $t>0$, from which 
the electric field 
$E(x, t) = {\rm Re}[ E(k, t) \exp ( i \; k \;  x ) ]$ is obtained. 
Hereafter, we represent the order of magnitude of the small perturbation amplitude by 
$\alpha$.  
Then, the linear solutions 
$f_1(x, v, t) = {\rm Re} [f_1 (k, v, t) \exp ( i \; k \;  x )] = {\cal O}(\alpha)$ 
and 
$E(x, t) = {\rm Re} [E(k, t) \exp ( i \; k \;  x )] = {\cal O}(\alpha)$ 
are used to derive 
\begin{eqnarray}
\langle f \rangle (v, t) a (v, t) 
& = & 
- \frac{e}{m} 
\langle E(x, t) f(x, v, t) \rangle 
\nonumber \\ 
& = &
- \frac{e}{2m} {\rm Re} 
[ E^* (k, t) f_1 (k, v, t) ]
,
\end{eqnarray}
where higher order terms than ${\cal O}(\alpha^2)$ are neglected. 
Now, we can write 
\begin{equation}
\langle f \rangle (v, t) 
=
f_0 (v) + f_2 (v, t)
,
\end{equation}
where we keep terms only up to ${\cal O}(\alpha^2)$ and $f_2 (v, t)$ 
represents the ${\cal O}(\alpha^2)$ part given by 
\begin{eqnarray}
& & 
 f_2 (v, t)
= \int_0^t dt' \frac{\partial}{\partial v} \left[
\frac{e}{m} 
\langle E(x, t') f(x, v, t') \rangle 
\right]
\nonumber \\ & & 
= \frac{e}{2m}
 \int_0^t dt' 
\;
{\rm Re} 
\left[ 
E^* (k, t') \frac{\partial f_1 (k, v, t')}{\partial v} 
\right]
.
\end{eqnarray}
We can also write 
\begin{equation}
\langle f \rangle (v, t) a (v, t) 
=
f_0  (v) a (v, t) + {\cal O}(\alpha^4)
,
\end{equation}
and neglect the ${\cal O}(\alpha^4)$ part to obtain 
\begin{equation}
a (v, t) 
=
- \frac{e}{2m} {\rm Re} 
\left[ 
E^* (k, t) \frac{f_1 (k, v, t)}{f_0(v)} 
\right]
,
\end{equation}
and
\begin{eqnarray}
\label{f2}
& & 
 f_2 (v, t)
=
-  \int_0^t dt' \frac{\partial}{\partial v} 
[ f_0(v) a(v,t)] 
\nonumber \\ & & 
=   f_0(v) 
  \int_0^t dt' 
\left[
\frac{m v}{T} a (v, t') - \frac{\partial a(v, t')}{\partial v} 
\right]
\nonumber \\ & & 
= f_0(v) 
\left[
\frac{\Delta {\cal E} (v, t)}{T}  - \Delta S(v, t)
\right]
,
\end{eqnarray}
where
\begin{eqnarray}
\label{Devt}
& & 
\Delta {\cal E} (v, t) 
\equiv
\int_0^t dt' 
\; m \,  v \, a (v, t')
\nonumber \\ & & 
=
- \frac{e}{2} v
\int_0^t dt' \;
{\rm Re} 
\left[
E^* (k, t') \frac{f_1 (k, v, t')}{f_0(v)} 
\right]
,
\end{eqnarray}
and 
\begin{eqnarray}
\label{DSvt}
& &
\Delta S(v, t) 
\equiv 
S(v, t) - S(v, 0)
=
\int_0^t dt' \;
\frac{\partial a(v, t')}{\partial v}
\nonumber \\ & & 
= 
- \frac{e}{2m}
\int_0^t dt' \;
 {\rm Re} 
\left[ 
E^* (k, t') \frac{\partial}{\partial v}
\left( \frac{f_1 (k, v, t')}{f_0(v)}  \right)
\right]
.
\end{eqnarray}
Since $a(v, t) = {\cal O}(\alpha^2)$, 
the variation in the electron velocity during the time interval $[0, t]$ 
is of ${\cal O}(\alpha^2)$. 
Neglecting ${\cal O}(\alpha^4)$ effects, 
we can regard $\Delta {\cal E} (v, t)$ as 
the change in the kinetic energy  
of the electron with the initial velocity $v$ caused by the acceleration due to the electric field 
during the time interval $[0, t]$. 
Recall that $\partial a(v, t)/\partial v$ represents the rate of change in 
the information content $S(v,t) \equiv -  \log \langle f \rangle (v, t)$ associated with the 
distribution of electrons in the velocity space. 
Then, $\Delta S(v,t)$ represents the change in the information content  
(or the information entropy) of the electron with the initial velocity $v$) 
during the time interval $[0, t]$. 

From Eq.~(\ref{f2}), we find 
\begin{equation}
\int_{-\infty}^{+\infty} dv \; 
 f _2 (v, t)
= 
\int_{-\infty}^{+\infty} dv \; 
f_0(v) 
\left[
\frac{\Delta {\cal E} (v, t)}{T}  - \Delta S(v, t)
\right]
= 0
\end{equation}
and 
\begin{eqnarray}
\label{DeltaE}
& & 
\frac{\Delta {\cal E} (t)}{T}
\equiv
\frac{1}{T}
\int_{-\infty}^{+\infty} dv \; 
f_0(v) 
\; \Delta {\cal E} (v, t) 
\nonumber \\ & & 
= 
\int_{-\infty}^{+\infty} dv \; 
f_0(v) 
 \; \Delta S(v, t)
\equiv \Delta S[\langle f \rangle ] (t)
\end{eqnarray}
which implies that, to the second order in $\alpha$, 
the change in the information (or Gibbs) entropy 
$\Delta S[\langle f \rangle ] (t) \equiv S[\langle f \rangle ] (t) - S[\langle f \rangle ] (0)$ 
is equal to the inverse temperature $1/T$ multiplied by 
the energy transfer $\Delta {\cal E} (t)$ from the electric field energy to 
the kinetic energy during the time interval $[0, t]$. 

\subsection{Calculation of $\Delta {\cal E} (v, t)$, $\Delta S(v,t)$, and $f_2(v,t)$}

Neglecting ${\cal O}(\alpha^4)$ effects, 
the change $\Delta {\cal E} (v, t)$ in the kinetic energy and the change  $\Delta S(v, t)$ in the information content  
(or the information entropy) of the electron with the initial velocity $v$ during the time interval $[0, t]$ 
are given by Eqs.~(\ref{Devt}) and (\ref{DSvt}), 
respectively. 
The electric field $E(k, t)$ and the perturbed distribution function $f_1(k, v, t)$ appearing 
in Eqs.~(\ref{Devt}) and (\ref{DSvt}) are evaluated 
using Eqs.~(\ref{solE}) and (\ref{solf1}), respectively, 
in which the series summations involving an infinite number of 
the complex-valued eigenfrequencies $\{ \omega_\mu \}$ 
converge more quickly for larger time $t$. 
On the other hand, 
the Taylor expansions of $E(k, t)$ and $f_1(k, v, t)$ about $\tau = k v_T t = 0$ 
are given by Eqs.~(\ref{EtTau}) and (\ref{f1Tau}), respectively,  
which converge more quickly for smaller time $t$. 
Therefore, to evaluate $\Delta {\cal E} (v, t)$ and $\Delta S(v,t)$, 
the expressions of $E(k, t)$ and $f_1(k, v, t)$ in Eqs.~(\ref{EtTau}) and (\ref{f1Tau}) 
and those in Eqs.~(\ref{solE}) and (\ref{solf1}) are separately 
used for smaller and larger time, respectively, and truncation to finite terms 
is made in these expressions 
so that the time integrals can be performed not numerically 
but analytically because 
the time dependence of the integrands are given in the form of the summation of 
exponential functions multiplied by polynomials of time. 
Under the conditions of examples shown later, 
good convergence is verified when 
including terms up to ${\cal O}(\tau^{13})$ in the integrands 
in Eqs.~(\ref{Devt}) and (\ref{DSvt}) 
for $\tau \equiv k v_T t \leq 1$  and 
using two pairs of the complex frequencies 
for Eqs.~(\ref{solE}) and (\ref{solf1}) to obtain 
$E(k, t)$ and $f_1(k, v, t)$
for $\tau \equiv k v_T t \geq 1$.
Once that $\Delta {\cal E} (v, t)$ and $\Delta s_V(v, t)$ are obtained following the 
procedures described above, 
we can use Eq.~(\ref{f2}) to calculate $f_2 (v, t)$ by
\begin{equation}
\label{f2Omegat}
 f_2 (v, t)
  = 
f_0(v) \; 
\Omega_t (v)
, 
\end{equation}
where 
\begin{equation}
\label{Omegatv}
\Omega_t (v)
\equiv
\frac{\Delta {\cal E} (v, t)}{T}  - \Delta S(v, t)
.
\end{equation}

\begin{figure}[h]
\centering
\includegraphics[width=8cm]{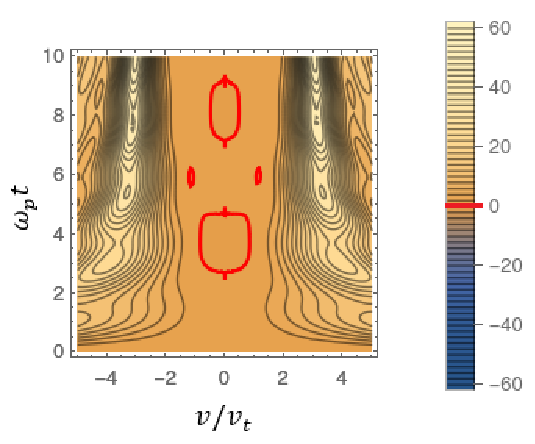}
\includegraphics[width=8cm]{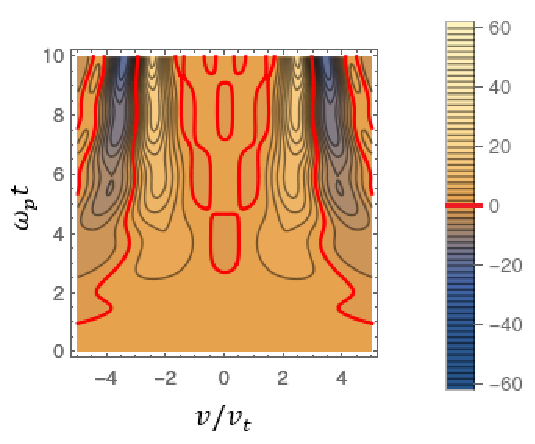}
\caption{
Contours of 
$\Delta {\cal E}(v, t)/( \alpha^2 T )$ (top) and $\Delta S(v, t)/\alpha^2$ (bottom) 
on the $(v, t)$ plane for $k \lambda_D = 1/2$.
}
\label{fig:devt}
\end{figure}

Figure~\ref{fig:devt} shows 
contours of $\Delta{\cal E}(v, t) / T$ and $\Delta S(v, t)$ on the $(v, t)$ plane, 
obtained from the analytical solution using the aforementioned procedure for $k \lambda_D = 1/2$. 
In most of the $(v,t)$ plane, we find $\Delta {\cal E} (v, t)  > 0$ which indicates
the increase of the electron kinetic energy. 
Here, the resonance velocity is given by 
$v_{res} \equiv  {\rm Re}(\omega_{\mbox{\o}}) / k  =  2.831 \;  v_t$.  
As time progresses, 
$\Delta {\cal E} (v, t)$ becomes more concentrated around 
the resonance velocities $v = \pm v_{res}$ while 
$\Delta S(v, t)$ clearly shows the positive maximum 
(negative minimum)  
at a slightly smaller (larger) absolute value $|v|$ than $v_{res}$. 

\begin{figure}[h]
\centering
\includegraphics[width=8cm]{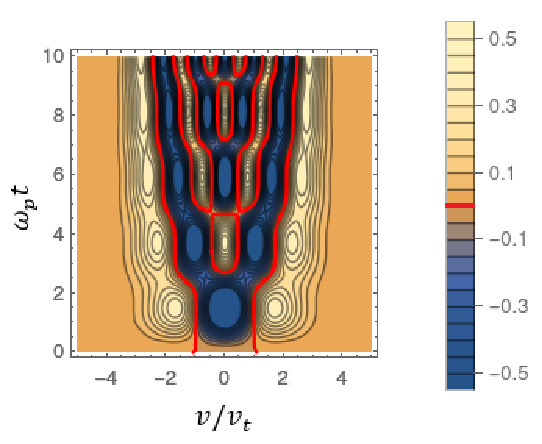}
\includegraphics[width=8cm]{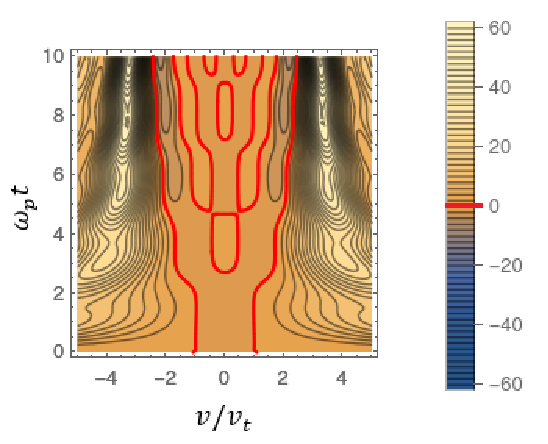}
\caption{
Contours of $f_2(v,t)/( \alpha^2 n_0 v_t^{-3} )$ (top) and $f_2(v,t)/( \alpha^2 f_0 (v) )$ (bottom) 
on the $(v, t)$ plane for $k \lambda_D = 1/2$. 
}
\label{fig:f2111315}
\end{figure}

\begin{figure*}[ht]
\centering
\includegraphics[width=18cm]{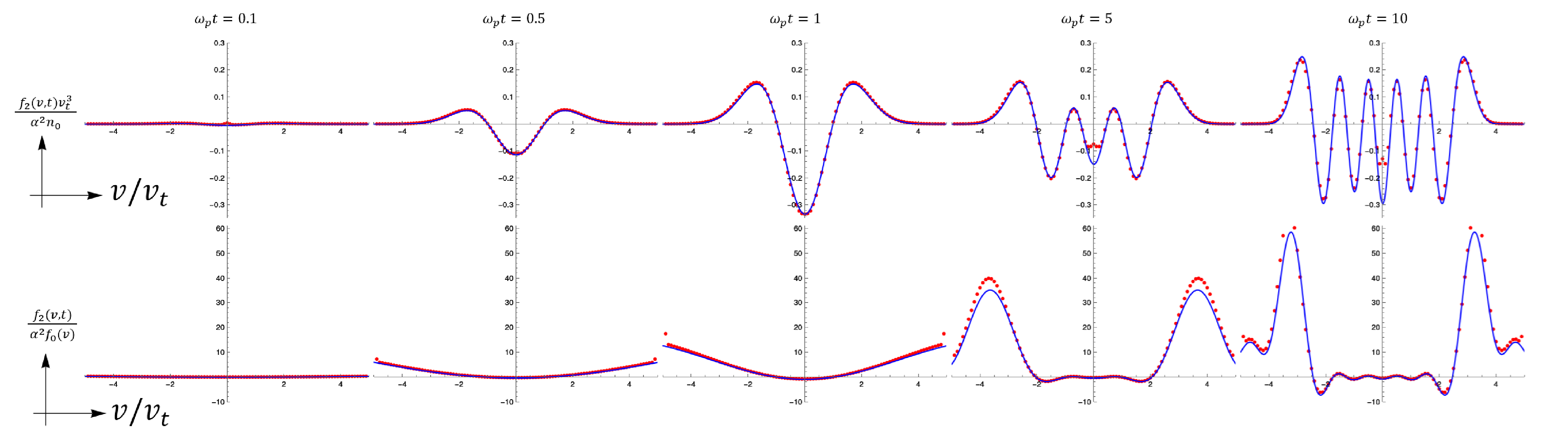}•
\caption{
Profiles of $f_2 (v, t) / ( \alpha^2 n_0 v_t^{-3} )$ (top) and $f_2(v, t) / (\alpha^2 f_0(v) )$ (bottom) 
 for $\omega_p t = 0.1$, 0.5, 1, 5, and 10. 
Results obtained by the analytical formulas and the CD simulation for $\alpha =0.1$ 
 are shown by \diffColor curves and red dots. 
 }
 \label{fig:f2-f2f0}
\end{figure*}

The top and bottom panels of Fig.~\ref{fig:f2111315} respectively show the distributions of 
$f_2(v,t)/(\alpha^2 n_0 v_t^{-3})$ and $f_2(v,t)/(\alpha^2 f_0(v))$ in the $(v, t)$ plane, 
calculated from the difference between $\Delta{\cal E}(v, t) / T$ and $\Delta S(v, t)$ using Eqs.~(\ref{f2Omegat}) and (\ref{Omegatv}).
As time progresses, the distribution of $f_2(v,t)$ spreads from around $v = 0$ to $v_{res}$ 
while such spreading is not clearly seen in the distribution of $f_2(v,t)/f_0(v)$. 
Figure~\ref{fig:f2-f2f0} shows $f_2(v,t)/(\alpha^2 n_0 v_t^{-3})$ and $f_2(v,t)/(\alpha^2 f_0(v,t))$ 
as functions of $v$, obtained from the analytical solution for $\omega_p t = 0.1, 0.5, 1, 5$, and 10. 
We see that  $f_2(v,t)$ oscillate along the $v$ direction and 
the number of oscillations increases with increasing time. 
Positive peaks and negative troughs of $f_2(v,t)/f_0(v)$ can be seen 
around the resonant velocities $v = \pm v_{res}$. 
The red dots represent the results of CD simulations for $\alpha=0.1$, which agree well with the analytical solution. 
However, for $\omega_p t \geq 5$, a significant discrepancy between the CD simulation results 
and the analytical solution appears near $v=0$. 
This discrepancy is attributed to the fact that the distribution function 
treated in the CD simulations is flat between a finite number of contour lines.
This results in an underestimation of the amplitude of the perturbed distribution function $f_1(k,v,t)$  
driven by $\partial f_0(v)/\partial v$ which is set to zero in the neighborhood of $v=0$ 
for the CD simulations. 
Then, the absolute value of $f_2(v,t)$ around $v=0$ is also underestimated. 
Indeed, it is confirmed that increasing the number of contour lines and narrowing the intervals 
between them in CD simulations reduces the difference between the CD simulation results and 
the analytical solution near $v=0$.

\begin{figure*}[ht]
\centering
\includegraphics[width=16cm]{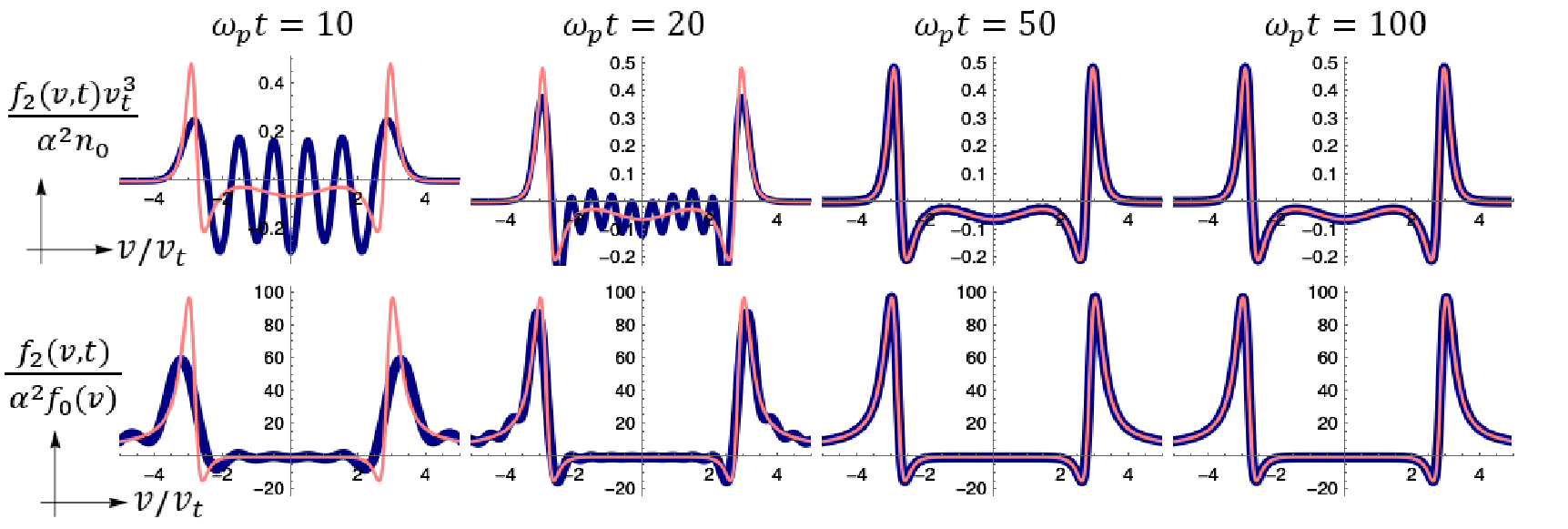}
\caption{
Profiles of $f_2 (v, t) /( \alpha^2 n_0 v_t^{-3} )$ (top) and $f_2(v, t) / (\alpha^2 f_0(v) )$ (bottom) 
obtained by the analytical formulas for $\omega_p t = 10$, 20, 50, and 100 
are shown by \diffColor curves. 
The red curves represent the profile in the limit of $t \rightarrow + \infty$. 
}
\label{f2infty}
\end{figure*}

Profiles of $f_2 (v, t) /( \alpha^2 n_0 v_t^{-3} )$ and $f_2(v, t) / (\alpha^2 f_0(v) )$ 
obtained by the analytical formulas for $\omega_p t = 10$, 20, 50, and 100
are shown by \diffColor curves in the top and bottom panels of Fig.~\ref{f2infty}, respectively.  
The red curves in Fig.~\ref{f2infty} represent the profiles in the long-time limit. 
As  $t \rightarrow + \infty$,  $f_2 (v, t)$ and $f_2 (v, t)/f_0(v)$  
converge to the structures which have the positive maximums (negative minimums) 
at $|v|$ slightly larger (smaller) than $v_{res} =  2.831 \;  v_t$. 
This is consistent with the well-known picture of the increase and decrease in particles' number around 
the resonant velocity in the Landau damping process.


\begin{figure}[h]
\centering
\includegraphics[width=8cm]{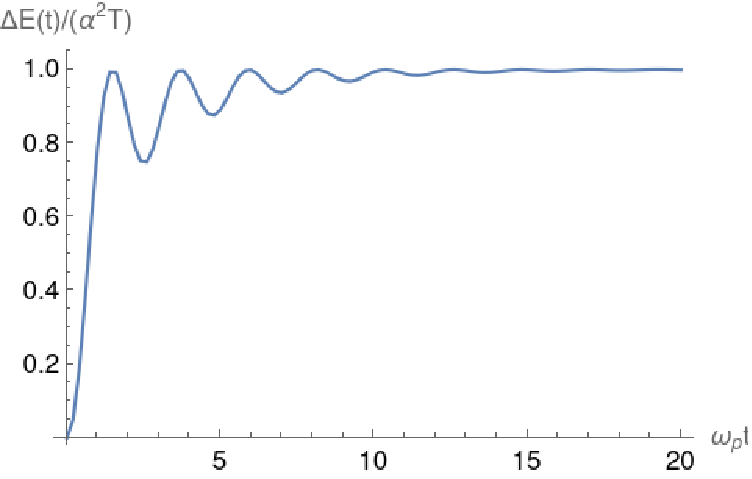}
\includegraphics[width=8cm]{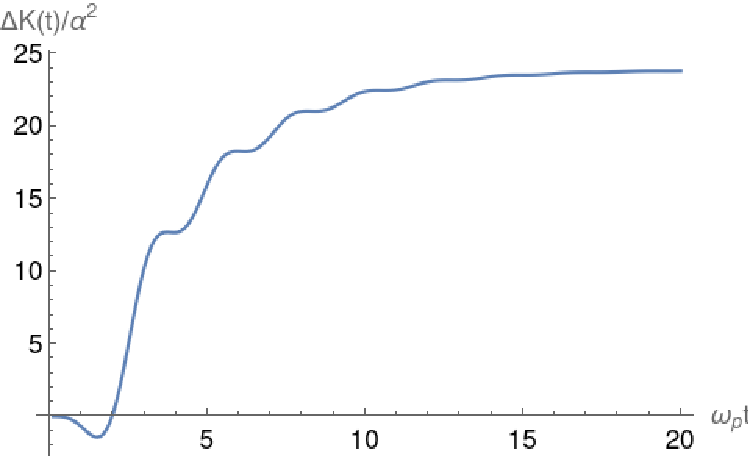}
\caption{
Time evolutions of 
$\Delta {\cal E}(t)/(\alpha^2 T)$ (top) and
$\Delta K(t)/\alpha^2$ (bottom).
}
\label{fig:DeltaE}
\end{figure}

For an arbitrary function $A(v)$, the change in its average value 
during $[0,t]$ is evaluated using Eqs.(\ref{f2Omegat}) and (\ref{Omegatv}) 
as
\begin{equation}
\label{DeltaA}
\Delta \overline{A} (t)
= \frac{1}{n_0}
\int_{-\infty}^{+\infty} dv \; 
 f_2 (v, t) A(v)
 = 
 \frac{1}{n_0}
\int_{-\infty}^{+\infty} dv \; 
 f_0(v) \; 
\Omega_t (v)
A(v)
. 
\end{equation}
By setting $A(v) = m v^2 / 2$, it can be shown that the above equation is equal to the increase in kinetic energy $\Delta {\cal E}(t)$ over the time interval $[0,t]$ given by Eq.~(\ref{DeltaE}). Furthermore, by calculating Eq.~(\ref{DeltaA}) for $A(v) = v^2$ and $v^4$,
the kurtosis of the velocity as a random variable at time $t$ is given by 
\begin{eqnarray}
K(t)
& = & 
\frac{\overline{v^4} (t)}{
[\overline{v^2} (t)]^2 } 
=
\frac{\overline{v^4} (0) + \Delta \overline{v^4} (t)}{
[\overline{v^2} (0) + \Delta \overline{v^2} (t)]^2 } 
\nonumber \\ 
& = & 
\frac{3 + \Delta \overline{v^4} (t)/ [ \overline{v^2} (0)]^2 }{
[1 + \Delta \overline{v^2} (t)/ \overline{v^2} (0) ]^2 }
= 3 + \Delta K(t)
,
\end{eqnarray}
where $\Delta K(t)$ is given up to the second order in $\alpha$ by 
\begin{equation}
\Delta K(t)
= 
\frac{\Delta \overline{v^4} (t)}{ 
[ \overline{v^2} (0)]^2}
- 6
\frac{\Delta \overline{v^2} (t)}{
 \overline{v^2} (0) }
 =
 \frac{m^2 \Delta \overline{v^4} (t)}{T^2}
- 6
\frac{m \; \Delta \overline{v^2} (t)}{T}
.
\end{equation}
Both $\Delta {\cal E}(t)$ and $\Delta K(t)$ are proportional to $\alpha^2$. 
Figure~\ref{fig:DeltaE} shows time evolutions of $\Delta {\cal E}(t)/(\alpha^2 T)$ 
and $\Delta K(t)/\alpha^2$ for $k \lambda_D = 1/2$. 
It is seen that 
$\Delta {\cal E}(t)/(\alpha^2 T)$ converges to $\Delta {\cal E}(\infty)/(\alpha^2 T) = 1/ (4 k^2 \lambda_D^2) = 1$ 
as $t \rightarrow +\infty$. 
The electric field is a standing wave of the form $\sin k x$, 
and its amplitude becomes zero twice during one period of plasma oscillation 
$2 \pi / {\rm Re} (\omega_{\mbox{\o}}) = 4.438 / \omega_p$. 
Considering that the sum of the kinetic energy of electrons and the energy of the electric field remains constant, 
it can be observed from the figure that $\Delta {\cal E}(t)/(\alpha^2 T)$ oscillates, 
reaching its maximum value, $\Delta {\cal E}(\infty)/(\alpha^2 T) = 1/ (4 k^2 \lambda_D^2) = 1$, 
twice during the period $2 \pi / {\rm Re} (\omega_{\mbox{\o}})$, and converges as $t \rightarrow +\infty$. 
As seen from Figs.~\ref{fig:f2-f2f0} and \ref{f2infty}, $\Delta K(t)$ is negative at early times like $\omega_p t < 2$,  
which reflects the fact that $f_2(v, t)$ is more localized near $v = 0$ for small $t$ as seen in 
Figs.~\ref{fig:f2111315} and \ref{fig:f2-f2f0}. 
As $t$ increases,  peaks of $f_2(v)$ near the resonant velocities $v = \pm v_{res}$ become more prominent, 
and $\Delta K(t)$ takes positive values and increases, 
which indicates a relative increase in the number of electrons with larger $|v|$ compared to a Gaussian distribution. 
We find that $\Delta K(t)$ converges to $\Delta K(\infty) = 23.92 \alpha^2$  as $t \rightarrow +\infty$. 

\section{Time evolution of  information entropies}

In this section, we consider the time evolution of information 
entropies in the Vlasov-Poisson system. 
For that purpose, we use the solution of the initial value problem in 
Sec.~II that gives the perturbed density and the electric field of 
${\cal O}(\alpha)$ as 
\begin{equation}
n_1 (x, t) = n_1 (t) \cos (k \; x), 
\hspace*{5mm}
E(x, t) 
= E_1(t) \sin (k \; x)
,
\end{equation}
where 
\begin{equation}
E_1(t) 
= -  \frac{4 \pi e}{k} n_1(t)
.
\end{equation}
Here, $n_1(t)$ and $E_1(t)$ satisfy the initial conditions, 
\begin{equation}
n_1(t=0) = \alpha \; n_0
,
\hspace*{5mm}
E_1(t=0) 
= - \alpha \frac{4 \pi e n_0}{k} 
,
\end{equation}
and they approach zero as 
$t \rightarrow +\infty$, 
\begin{equation}
n_1 (x, \infty) = 0, 
\hspace*{5mm}
E(x,  \infty) = 0 
.
\end{equation}
The electric field energy density at time $t$ 
is given by 
\begin{eqnarray}
& & 
\frac{1}{8\pi}  
\langle E(x,t)^2 \rangle
= 
\frac{1}{8\pi}  \left ( \frac{4 \pi e}{k} \right)^2 
\langle [ n_1(x,t) ]^2 \rangle
\nonumber \\
&  & 
= 
\frac{1}{16\pi}  
[ E_1(t) ]^2 
= 
\frac{1}{16\pi}   \left ( \frac{4 \pi e}{k} \right)^2 
[ n_1(t) ]^2 
\nonumber \\
&  & 
=
\frac{n_0 T}{ 2 (k \lambda_D)^2}   
\frac{ \langle n_1(x,t)^2 \rangle}{n_0^2}
=
\frac{n_0 T}{ 4 (k \lambda_D)^2}   
\frac{[n_1(t)]^2}{n_0^2}
.
\end{eqnarray}
Also, from the energy conservation law, we obtain 
\begin{eqnarray}
\label{energy-conservation}
& & 
n_0 \Delta {\cal E} ( t ) 
+ \frac{1}{8\pi}  
\langle E(x,t)^2 \rangle
= 
n_0 \Delta {\cal E} ( \infty) 
\nonumber \\  
&  & 
=
\frac{1}{8\pi}  
\langle [ E(x,0) ]^2 \rangle
= 
\alpha^2  \frac{n_0 T}{4 ( k \lambda_D)^2} 
,
\end{eqnarray}
where $\Delta {\cal E} ( t )$ represents the change in the kinetic energy of the electron 
during the time interval $[0,t]$ 
given by Eq.~(\ref{DeltaE}).

\begin{figure}[h]
\centering
\includegraphics[width=8cm]{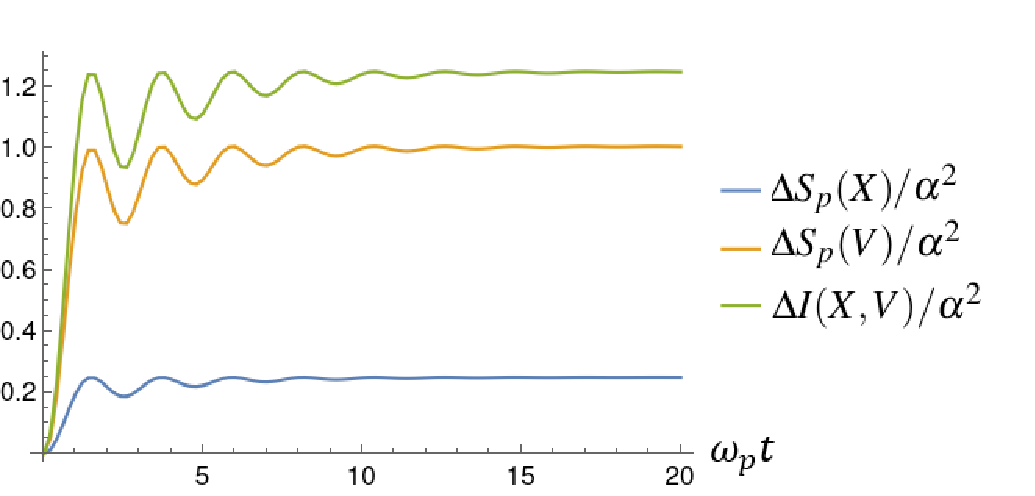}
\caption{
Time evolutions of 
$\Delta S_p(X) = (k \lambda_D)^2 \Delta{\cal E}(t) / (\alpha^2 T)$, 
$\Delta S_p(V) / \alpha^2 = \Delta {\cal E}(t) / (\alpha^2 T)$, and 
$\Delta I(X, V) /\alpha^2 = (\Delta S_p(X) + \Delta S_p(V) )/\alpha^2$.
}
\label{SXV}
\end{figure}

We regard the electron's position and velocity as random variables which are 
represented by $X$ and $V$, respectively, as explained in Appendix~E.  
The joint probability density function of $X$ and $V$ are denoted by 
$p(x, v, t)$ [see Eq.~(\ref{pxv})], which is integrated with respect to $x$ and $y$ to 
give the marginal probability density functions 
$p_X(x, t)$ and $p_V(v, t)$, respectively [see Eqs.~(\ref{px}) and (\ref{pv})]. 
The entropy $S_p(X)  \equiv S [p_X]$ is derived from the electron's position 
distribution function $p_X(x, t)$ as 
\begin{eqnarray}
\label{SpX}
& & 
S_p(X)  \equiv S [p_X] 
\equiv 
- \int_{-L/2}^{+L/2} dx  \; p_X(x, t) \log p_X(x, t)
\nonumber \\
& & 
=
-
  \int_{-L/2}^{+L/2} \frac{dx}{L}  \; \left(   \frac{n (x, t)}{n_0}  \right) 
\log  \left(   \frac{n  (x, t)}{n_0}  \right) 
+ \log ( L )
\nonumber \\
& & 
\simeq  \log ( L  )
- \frac{1}{2}
 \int_{-L/2}^{+L/2} \frac{dx}{L}
\left(  \frac{n_1  (x, t)}{n_0}  \right)^2  
.
\end{eqnarray}
In the last line of Eq.~(\ref{SpX}), 
terms of higher orders than $\alpha^2$ are neglected. 
The increase in $S_p(X)  \equiv S [p_X] $ during 
the time interval from $0$ to $t$ is given by 
\begin{eqnarray}
& & 
\Delta S [p_X] (t)
= 
S [p_X] (t) - S [p_X] (0)
\nonumber \\ & & 
\simeq 
\frac{1}{2}
 \int_{-L/2}^{+L/2} \frac{dx}{L}
 \left[
\left(  \frac{n_1  (x, 0)}{n_0}  \right)^2  
- \left(  \frac{n_1  (x, t)}{n_0}  \right)^2  
\right]
\nonumber \\ & & 
=
\frac{
\langle [ n_1  (x, 0) ]^2 -  [n_1  (x, t) ]^2 \rangle
}{ 2 n_0^2 }
.  
\end{eqnarray}
As $t \rightarrow \infty$, it approaches to 
\begin{equation}
\label{DSpX}
\Delta S [p_X] (\infty)
=
\frac{
\langle [ n_1  (x, 0) ]^2 \rangle
}{ 2 n_0^2 }
= \frac{\alpha^2}{4}
.
\end{equation}
The velocity distribution function $p_V(v, t)$ 
is used to express the 
entropy $S_p(V)  \equiv S [p_V]$ as 
\begin{eqnarray}
& & 
S_p(V)  \equiv S [p_V] 
\equiv 
- \int_{-\infty}^{+\infty} dv \; p_V(v, t) \log p_V(v, t)
\nonumber \\
& & 
=
- \frac{1}{n_0}
  \int_{-\infty}^{+\infty}  dv \; \langle f \rangle (v, t) 
\log   \langle f \rangle (v, t) 
+ \log ( n_0 )
.
\end{eqnarray}
The increase in $S_p(V)  \equiv S [p_V]$ during 
the time interval from $0$ to $t$ is represented by
\begin{eqnarray}
\label{DeltaSpV}
& & 
\Delta S [p_V] (t)
= 
S [p_V] (t) - S [p_V] (0)
\nonumber \\ & & 
\simeq 
\frac{1}{n_0}
  \int_{-\infty}^{+\infty}  dv \; f_2 (v, t) 
\frac{m v^2}{2 T}
= 
\frac{\Delta {\cal E} (t)}{T}
,
\end{eqnarray}
where terms of higher orders than $\alpha^2$ are neglected again. 
Using Eqs.~(\ref{DeltaSpV}) and 
(\ref{energy-conservation}), 
$\Delta S [p_V] (t)$ can be rewritten as 
\begin{eqnarray}
\Delta S [p_V] (t)
& = &
\frac{\Delta {\cal E} (t)}{T}
=
\frac{1}{8\pi n_0 T} \left\langle
[ E(x, 0) ]^2 - [ E(x, t) ]^2 
\right\rangle
\nonumber \\
& = & 
\frac{1}{ 2 ( k \lambda_D)^2} 
\frac{
\left\langle
[ n_1(x, 0) ]^2 - [ n_1(x, t) ]^2 
\right\rangle
}{ n_0^2 }
. 
\end{eqnarray}
Its value in the limit of 
$t \rightarrow \infty$ is given by 
\begin{equation}
\label{DSpV}
\Delta S [p_V] (\infty)
= 
\frac{\Delta {\cal E} (\infty)}{T}
=   \frac{\alpha^2}{4 k^2 \lambda_D^2} 
. 
\end{equation}
The mutual information of the random variables
$X$ and $V$ is defined by 
\begin{equation}
I (X, V)
= S_p(X) + S_p(V) - S_p (X, V) 
.
\end{equation}
Under the initial condition given by Eq.~(\ref{initial}), 
$X$ and $V$ are statistically independent at $t=0$, 
and accordingly 
\begin{equation}
[ I (X, V) ]_{t=0}
= 0
. 
\end{equation}
When $f (x, v, t) = (n_0 L) p (x, v, t)$ satisfies the Vlasov equation in 
Eq.~(\ref{Vlasov1}), 
the entropy defined by
\begin{equation}
S_P (X, V) \equiv S[p] \equiv 
- \int_{-L/2}^{+L/2} dx \int_{-\infty}^{+\infty} dv \; p(x, v) \log p(x, v)
\end{equation}
is known to be an invariant, 
and 
the increase in $I (X, V)$ during the time interval from $0$ to $t$ 
is written as 
\begin{eqnarray}
& & 
\Delta I (X, V) = I(X, V)
= \Delta S_p(X) + \Delta S_p(V) 
\nonumber \\ & & 
= 
\frac{
\langle
[ n_1  (x, 0) ]^2  
- [ n_1  (x, t) ]^2  
\rangle 
}{2n_0^2}
+ 
\frac{\Delta {\cal E} (t)}{T}
\nonumber \\ & & 
= 
\frac{
\langle [ n_1  (x, 0) ]^2 -  [n_1  (x, t) ]^2
}{ 2 n_0^2 }
\left(
1 + \frac{1}{k^2 \lambda_D^2} 
\right)
\nonumber \\ & & 
= 
\frac{\Delta {\cal E}(t)}{T}
( 1 + k^2 \lambda_D^2 )
. 
\end{eqnarray}
In the limit of 
$t \rightarrow \infty$, we obtain 
\begin{equation}
[ \Delta I (X, V) ]_{t = \infty}
=
\frac{\alpha^2}{4}
\left(
1 + \frac{1}{k^2 \lambda_D^2} 
\right)
.
\end{equation}
The time evolutions of 
$\Delta S_p(X) \equiv \Delta S[p_X] = (k \lambda_D)^2 \Delta{\cal E}(t) / T$, 
$\Delta S_p(V) \equiv \Delta S[p_V] = \Delta {\cal E}(t) / T$, and 
$\Delta I(X, V) = \Delta S_p(X)+ \Delta S_p(V)$ are shown in 
Figure~\ref{SXV}, where contributions of higher orders than  
${\cal O}(\alpha^2)$ are neglected.  
As functions of time $t$, 
$\Delta S_p(X) \equiv \Delta S[p_X]$, $\Delta S_p(V)\equiv \Delta S[p_V]$ and $\Delta I(X, V)$ 
have the same form but different magnitudes.

\begin{figure}[h]
\centering
\includegraphics[width=8cm]{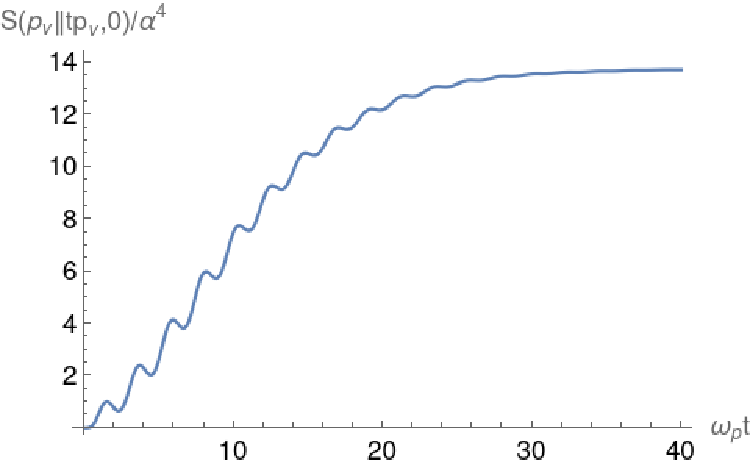}
\caption{
Time evolution of the relative entropy $S(p_V, t || p_V, 0)$ normalized by $\alpha^4$. 
}
\label{fig:relative_entropy}
\end{figure}

Comparing the velocity distribution functions 
$p_V(v,t) = \langle f \rangle (v, t) / n_0$ and 
$p_V(v,0) = \langle f \rangle (v, 0) / n_0$, 
and using Eqs.~(\ref{relative_entropy}) and (\ref{relative_entropy2}),  
we obtain the relative entropy $S(p_V, t || p_V, 0)$ as 
\begin{eqnarray}
S(p_V, t || p_V, 0) 
& = &
\int_{-\infty}^{+\infty} dv \;
p_V(v, t) \log \left[
\frac{p_V(v, t)}{p_V(v, 0)}
\right]
\nonumber \\
& \simeq &
\frac{1}{2 n_0}
\int_{-\infty}^{+\infty} dv \;
\frac{ [ f_2 (v, t)  ]^2}{
f_0  (v) }
\nonumber  \\ 
& = & 
\frac{1}{2 n_0}
\int_{-\infty}^{+\infty} dv \;
f_0  (v) 
[\Omega_t(v) ]^2
, 
\end{eqnarray}
which takes a small value of ${\cal O}(\alpha^4)$. 
Figure~\ref{fig:relative_entropy} shows 
the time evolution of $S(p_V, t || p_V, 0)$. 
As $t$ increases, $S(p_V, t || p_V, 0)$ increases while showing modulation 
and approaches to the limit value that corresponds to the limit function 
$\lim_{t \rightarrow +\infty} f_2(v, t)$ in Fig.~\ref{f2infty}. 
As described in Appendix~E, 
the non-negativity of
$S(p_V, t || p_V, 0)$ implies that,  
even in the collisionless Vlasov-Poisson system, 
the inequality in the form of the second law of thermodynamics
holds in the relation between the heat transfer from the Maxwellian velocity distribution 
to the electric field 
and the conditional entropy of the electron position variable for
a given velocity distribution. 

\section{Collisional relaxation to thermal equilibrium}

Coulomb collisions are neglected in the Vlasov equation~(\ref{Vlasov0}). 
However, even if the collision frequency is very small but finite,
Coulomb collisions eventually relax the distribution function to the Maxwellian 
\begin{equation}
f_{M\infty} = n_{M\infty} \left( \frac{m}{2\pi T_{M\infty}}  \right)^{1/2}
\exp \left[
- \frac{m}{2 T_{M\infty}} (v - u_{M\infty} )^2
\right]
\end{equation}
at  $t = + \infty$ while conserving total particles' number, momentum, and energy.
The Maxwellian distribution function 
$f_{M\infty}(v)$ is the equilibrium solution of the Boltzmann equation 
that is given by including the collision term in the Vlasov equation. 
When the collision operator acts on the Maxwellian, the collision term 
vanishes. 
Then, substituting $f_{M\infty}(v)$ into the left-hand side into the Vlasov equation, 
it also vanishes. 
It can be proven from the above-mentioned fact that 
$n_{M\infty}$, $T_{M\infty}$, and $u_{M\infty}$ are all independent 
of $t$ and $x$. 
Under the initial condition in Eq.~(\ref{initial}), 
we can use conservation laws for particles' number, energy, and momentum
to derive 
\begin{equation}
\label{nuTM}
n_{M\infty} = n_0,
\hspace*{5mm}
u_{M\infty} = 0,
\hspace*{5mm}
\frac{n_0 T_{M\infty}}{2} 
 = \frac{n_0 T_0}{2} 
 + \frac{\langle [ E(x,0) ]^2 \rangle}{8\pi} 
 .
\end{equation}
Thus, the distribution function in the thermal equilibrium state reached 
by the collisional relaxation is given by 
\begin{equation}
f_{M\infty} = n_0 \left( \frac{m}{2\pi T_{M\infty}}  \right)^{1/2}
\exp \left(
- \frac{m v^2}{2 T_{M\infty}} 
\right)
.
\end{equation}
Here, ions are treated as uniform background positive charge with infinite mass, 
so no energy exchange between electrons and ions 
due to collisions is considered to occur.
From Eq.~(\ref{nuTM}), we also have
\begin{equation}
\frac{\Delta T_{M\infty}}{T}
 = 
\frac{T_{M\infty} - T}{T}
 = 
 \frac{\langle [ E(x,0) ]^2 \rangle}{4\pi n_ 0 T} 
=
\frac{\alpha^2}{2(k \lambda)^2}
= 
2 \frac{\Delta {\cal E}(\infty)}{T}
.
\end{equation}

From the equilibrium probability distribution function 
$p_M (x,v) = f_{M\infty} (v) / ( n_0 L )$ and its marginal probability 
distribution functions $p_{MX} (x) = 1/L$ and 
$p_{MV}(v) = f_{M\infty}(v) / n_0$, 
we can define the entropies 
$S [p_M]$, $S[p_{MX}]$, and $S[p_{MV}]$ 
in the thermal equilibrium 
where
\begin{equation}
p_M = p_{MX} \cdot p_{MV}
\end{equation}
holds so that the random variables $X$ and $V$ are statistically 
independent, and accordingly, the mutual information 
of $X$ and $V$ vanishes, 
\begin{equation}
I_M (X, V)
\equiv
S [p_M] - S[p_{MX}] - S[p_{MV}]
 = 
0
.
\end{equation}
The deviation of the entropy $S[p_{MX}]$ from the initial value $S[p_X](t=0)$ 
is given by 
\begin{eqnarray}
\label{DSpMX}
& & \Delta S [p_{MX}]
 \equiv 
S [p_{MX}] - S[p_X](t=0)
\nonumber \\
 & & = 
\left\langle
 [ 1+ \alpha \cos (kx) ]
\log  [ 1+ \alpha \cos (kx) ]
\right\rangle
\simeq
\frac{\alpha^2}{4}
.
\end{eqnarray}
We see from 
Eqs.~(\ref{DSpX}) and (\ref{DSpMX})
that $\Delta S [p_{MX}]$ and $\Delta S [p_X](\infty)$
agree with each other up to ${\cal O}(\alpha^2)$. 
Next, the deviation of $S[p_{MV}]$ from $S[p_V](t=0)$ 
\begin{eqnarray}
\label{DSpMV}
& & \Delta S [p_{MV}]
 \equiv 
S [p_{MV}] - S[p_V](t=0)
\nonumber \\
 & & = 
\frac{1}{2} ( \log T_{M\infty} - \log T)
\simeq \frac{\Delta T_{M\infty}}{2T} 
= \frac{\Delta {\cal E}(\infty)}{T}
.
\end{eqnarray}
It is also found that 
$\Delta S [p_{MV}] = \Delta S [p_V](\infty)$
up to ${\cal O}(\alpha^2)$.
The deviation of the entropy
$S[p_M]$ from its initial value $S[p](t=0)$ 
evaluated as
\begin{eqnarray}
 \Delta S [p_M]
 & \equiv & 
S [p_M] - S[p](t=0)
 = 
 \Delta S [p_{MX}]  + \Delta S [p_{MV}]
 \nonumber \\
 & \simeq & 
\frac{\alpha^2 }{4}
+\frac{\Delta {\cal E}(\infty)}{T}
=
\frac{\alpha^2 }{4}
\left ( 
1+ \frac{1}{k^2 \lambda_D^2}
\right)
 \nonumber \\
& = & 
\frac{\Delta {\cal E}(\infty)}{T}
\left ( 
1+ k^2 \lambda_D^2
\right)
.
\end{eqnarray}
We note that 
$S[p]$ is invariant in the collisionless process although
it increases by $\Delta S [p_M]$ when the system reaches the 
thermal equilibrium state due to collisions. 

\begin{figure*}[ht]
\centering
\includegraphics[width=16cm]{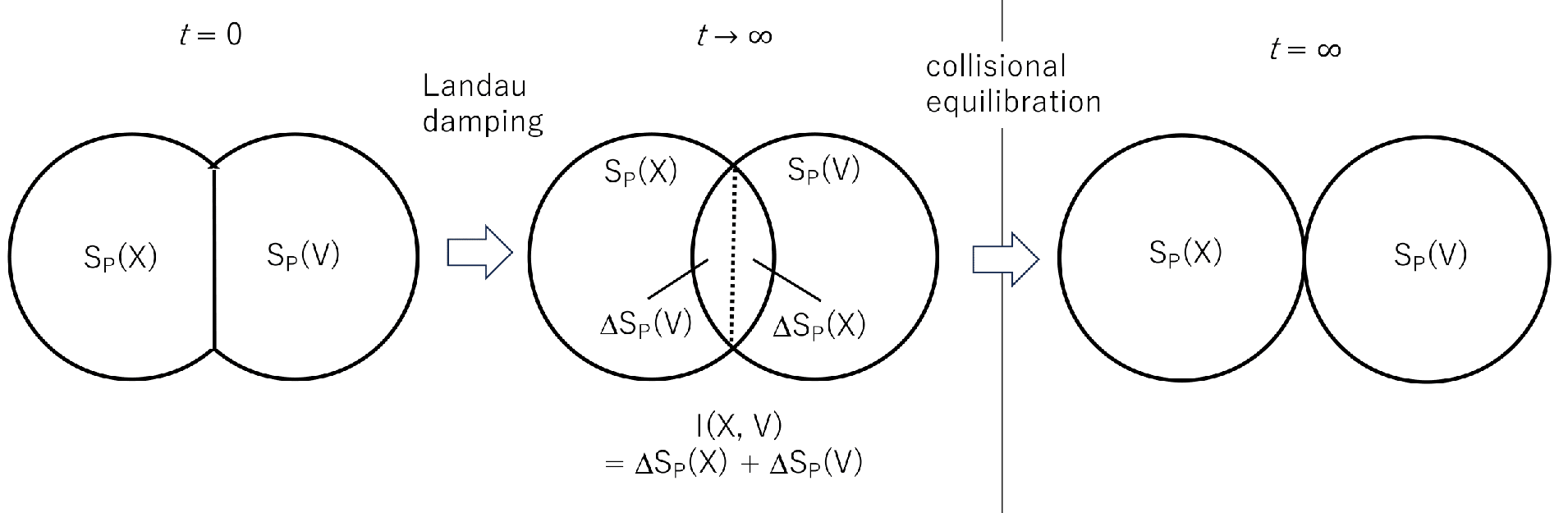}
\caption{
The entropies $S_P(X)$, $S_P(V)$, $S_P(X,V)$ and the mutual information $I(X,V)$
for $t=0$ (left), $t \rightarrow + \infty$ in the collisionless process (middle), and 
$t = + \infty$ in the collisional process (right). 
Their magnitudes are represented by the areas inside the corresponding contours.
}
\label{fig:entropy}
\end{figure*}

In Fig.~\ref{fig:entropy} , the magnitudes of the entropies $S_P(X)$, $S_P(V)$, $S_P(X,V)$, 
and the mutual information $I(X,V)$ are 
represented by the area inside the corresponding contours for $t=0$, 
$t \rightarrow +\infty$ in the collisionless process, and 
$t = +\infty$ in the collisional process. 
Here, the entropies $S_P(X) \equiv S[P_X]$, $S_P(V) \equiv S[P_X]$, and $S_P(X,V) \equiv S[P]$ 
takes non-negative values 
and they are related to $S_p(X) \equiv S[p_X]$, $S_p(V) \equiv S[p_V]$, 
and $S_p(X,V) \equiv S[p]$ 
by the relations shown in Eqs.~(\ref{SPSp}) and (\ref{SPSpXV}). 
The entropy $S_P(X,V)$  does not change in the collisionless process 
although the Landau damping increases $S_P(X)$ and $S_P(V)$ 
by $\Delta S_P(X) = \alpha^2 / 4$ and 
$\Delta S_P(V) = \alpha^2 / (4 k^2 \lambda_D^2)$, respectively, 
as shown in Eqs.~(\ref{DSpX}) and (\ref{DSpV}), 
and the mutual information content $I(X,V)$ increases by 
$\Delta I(X,V)= \Delta S_P(X) +\Delta S_P(V)$.
Let us compare the limit state at $t \rightarrow +\infty$ in the collisionless process 
and the thermal equilibrium state reached by collisions. 
In the latter state, the values of $S_P(X)$ and $S_P(V)$ 
remain the same as in the former up to the ${\cal O}(\alpha^2)$ accuracy. 
However, in the thermal equilibrium,
the mutual information quantity $I(X,V)$ vanishes and 
the entropy $S_P(X,V)$ of the whole system increases 
by the amount that $I(X,V)$ decreases.

\section{Conclusions and Discussion}

In this paper, 
the one-dimensional Vlasov-Poisson system describing a plasma consisting of electrons and uniformly distributed ions 
with infinite mass is considered. 
Using analytical solutions and contour dynamics simulations, 
we elucidate how the information entropies determined from the distribution functions of the electron position and velocity variables evolve 
in the Landau damping process.
Under the initial condition given by the Maxwellian velocity distribution with the perturbed density distribution in the form of the cosine function, 
linear and quasilinear analytical solutions describing the time evolutions of the electric field and the distribution function 
are obtained and shown to be in good agreement with results from numerical simulations based on contour dynamics.

A novel approximate integral formula including the effect of an infinite number of complex eigenfrequencies to correctly evaluate the electric field is presented. 
In addition, the linear analytical solutions for the electric field and the distribution function 
near the initial time are expressed as series expansions in time and velocity variables.
The quasilinear analytical solution describing the time evolution of the spatially averaged velocity distribution function is 
obtained, and its validity is confirmed by the contour dynamics simulation results.
These analytical expressions of the linear and quasilinear solutions are useful 
for verification of the accuracy of simulations of the Vlasov-Poisson system 
using methods other than the contour dynamics as well. 
Using the quasilinear analytical solution, 
it becomes possible to accurately determine the time evolutions of the electron kinetic energy 
and the background velocity distribution function associated with the Landau damping. 
Furthermore, the time evolutions of the information entropies of the electron position 
and velocity variables, and the mutual information are 
determined with an accuracy of the order of the squared perturbation amplitude $\alpha^2$. 
It is well known that, in a collisionless process, the information entropy determined from the joint probability density distribution function of position and velocity variables (or the phase-space distribution function) is one of the Casimir invariants.
On the other hand, 
the decrease in the squared mean of spatial density fluctuations increases the information entropy of the position variable, 
 and the ratio of the increase in the electron kinetic energy to the temperature equals the increase in 
 the information entropy of the velocity variable to the order of $\alpha^2$. 
 The sum of these increases in the information entropies of the position and velocity variables 
 yields the mutual information that is initially zero.  

The relative entropy obtained by comparing the velocity distribution at time $t$ 
with the initial distribution is a positive quantity of order of $\alpha^4$. 
This leads to the fact that, even in the collisionless process, 
the inequality in the form of the second law of thermodynamics
holds in the relation between the heat transfer from the Maxwellian velocity distribution 
to the electric field 
and the conditional entropy of the electron position variable for
a given velocity distribution. 

When Coulomb collisions are taken into account, 
they relax the distribution function at $t \rightarrow + \infty$ in the collisionless process 
further to the thermal equilibrium state. 
In this relaxation, the mutual information of the position and velocity variables decreases to zero, although 
the information entropies of the position and velocity variables do not change to the order of $\alpha^2$. 
Then, the entropy determined from the phase-space distribution increases by the amount of the decrease in the mutual information. 
It indicates the validity of Boltzmann's H-theorem. 
Future extensions of the present work include studies on 
the position dependence of the phase-space distribution function of order $\alpha^2$, which is not included in the quasilinear solution, 
and the analysis of the information entropies and the mutual information of the position and velocity variables 
to the order of $\alpha^4$.

\begin{acknowledgments}
This work is supported in part by 
the JSPS Grants-in-Aid for Scientific Research (Grant Nos.~19H01879 and 24K07000)
 and in part by the NINS program of Promoting Research by Networking among Institutions 
 (Grant No.~01422301). 
 Simulations in this work were performed on ``Plasma Simulator'' (NEC SX-Aurora TSUBASA) of 
 NIFS with the support and under the auspices of the NIFS Collaboration Research program 
 (Grant Nos.~NIFS23KIPT009 and NIFS24KISM007).
\end{acknowledgments}

\section*{AUTHOR DECLARATIONS}

\subsection*{Conflict of Interest}

The authors have no conflicts of interest to disclose.

\subsection*{Author Contributions}
\noindent
{\bf K. Maekaku}: Data curation (lead); Investigation (lead); Visualization (lead); Software (lead);  Methodology (equal); Writing--original draft (equal); Writing--review \& editing (lead). 
{\bf H. Sugama}: Conceptualization (lead); Formal analysis (lead); Investigation (supporting); Methodology (equal); Writing--original draft (equal); Writing-review \& editing (supporting); Supervision (lead).
{\bf T.-H. Watanabe}: Methodology (equal); Writing--review \& editing (supporting).

\section*{DATA AVAILABILITY}
The data that support the findings of this study are available from the corresponding authors upon reasonable request.

\appendix

\section{Plasma dispersion function}

The plasma dispersion function is defined by~\cite{Stix,Miyamoto}
\begin{equation}
\label{Z}
Z(\zeta)
=  \frac{1}{\sqrt{\pi}}
\int_{-\infty}^{+\infty} dz \frac{e^{-z^2}}{z - \zeta}
,
\end{equation}
in the case of ${\rm Im} \zeta > 0$ .
Analytic continuation needs to be done to define $Z(\zeta)$ in the case of ${\rm Im} \zeta  \leq 0$. 
For ${\rm Im} \zeta  = 0$, we have
\begin{equation}
Z(\zeta)
=  \frac{1}{\sqrt{\pi}} P
\int_{-\infty}^{+\infty} dz \frac{e^{-z^2}}{z - \zeta}
+ i \sqrt{\pi} e^{-\zeta^2}
.
\end{equation}
The plasma dispersion function is also written as 
\begin{equation}
Z(\zeta)
= 
i  \sqrt{\pi} e^{-\zeta^2}
\left[
1 + {\rm erf} ( i \zeta )
\right]
,
\end{equation}
where the error function ${\rm erf}$ is defined by 
\begin{equation}
{\rm erf} \; z
= \frac{2}{\sqrt{\pi}} \int_0^z e^{-s^2} ds 
.
\end{equation}
The plasma dispersion function satisfies 
\begin{equation}
\label{Zcc}
 [ Z(-\zeta^*) ]^*
= 
-  Z(\zeta) 
,
\end{equation}
where $^*$ represents the complex conjugate. 
The derivative of $Z(\zeta)$ with respective to $\zeta$ is given by 
\begin{equation}
\label{dZ}
Z' (\zeta) \equiv \frac{dZ(\zeta)}{d\zeta}
= -2 [ 1 + \zeta Z(\zeta) ]
.
\end{equation}
The series expansion of $Z(\zeta)$ about $\zeta = 0$ is given by 
\begin{equation}
Z(\zeta)
=
i  \sqrt{\pi} e^{-\zeta^2}
- \sum_{n=1}^\infty
\frac{(-1)^{n-1} 2 \sqrt{\pi} }{\Gamma (n+ 1/2)} \zeta^{2n-1}
,
\end{equation}
and the asymptotic expansion of $Z(\zeta)$ for $|\zeta| \gg 1$ is written as
\begin{equation}
\label{asymptoticZ}
Z(\zeta)
=
- \sum_{n=1}^N
\frac{\Gamma (n - 1/2) }{\sqrt{\pi} } \frac{1}{\zeta^{2n-1}}
+ \frac{\Gamma (N + 1/2) }{\sqrt{\pi} } e^{-\zeta^2} 
\int_{i\infty}^\zeta \frac{e^{s^2}}{s^{2N}} ds 
.
\end{equation}
%

\color{\diffColor}
\section{Contours for integrals in the complex plane}
\begin{figure}[h]
\centering
\includegraphics[width=8cm]{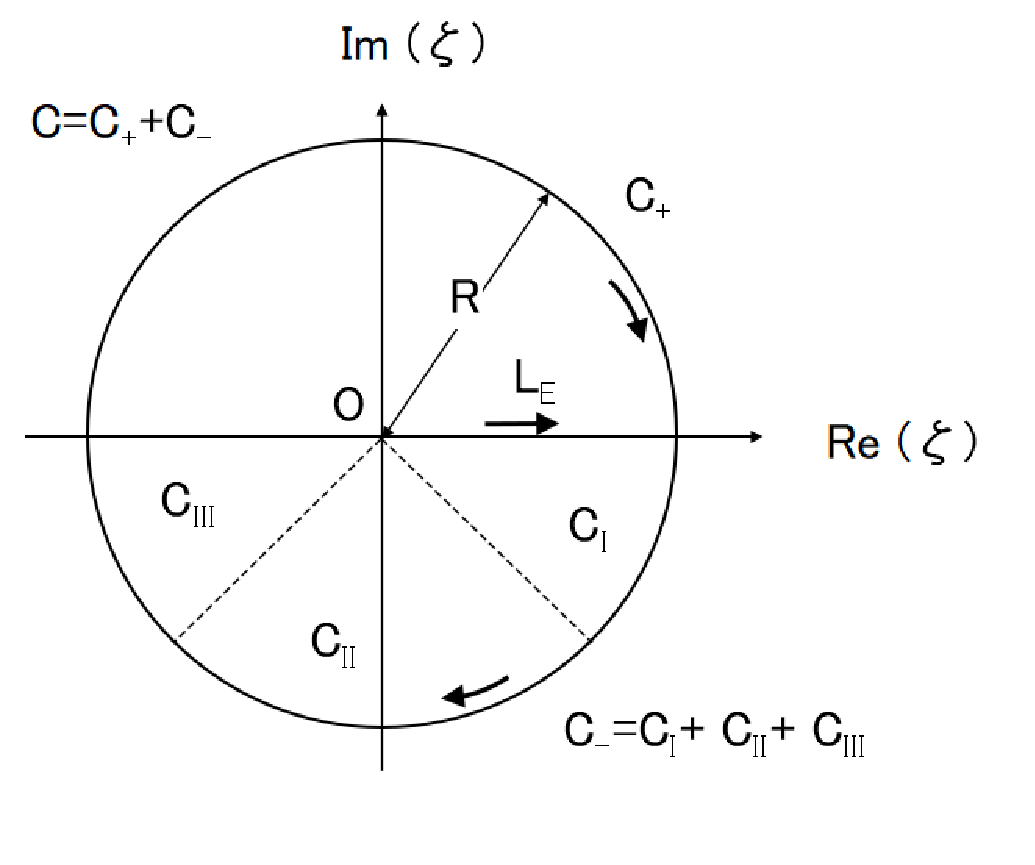}
\caption{
Contours $L_E$ and $C=C_+ + C_-$ in the complex $\zeta$-plane. 
Here, $L_E$ represents the real axis as the contour for integration in the complex $\zeta$-plane 
from ${\rm Re}(\zeta)= - \infty$ to ${\rm Re}(\zeta)= + \infty$. 
The circle $C$ with the center at the origin 0 and the radius $R \gg 1$ 
consists of the semicircles $C_+$ and $C_-$ which are 
in the upper $[{\rm Im}(\zeta)>0]$ and lower $[{\rm Im}(\zeta)<0]$ half-planes, respectively. 
The semicircle $C_- : \zeta = R e^{i \theta}$ is 
divided into the three arcs $C_{\rm I}$, $C_{\rm II}$, and $C_{\rm III}$ which 
correspond to $0 > \theta >  - \pi / 4$, $- \pi / 4 > \theta > - 3 \pi / 4$, and 
$- 3\pi / 4 > \theta > - \pi$, respectively. 
In the contour integrals, the orientations of $C$, $C_+$, $C_-$, 
$C_{\rm I}$, $C_{\rm II}$, and $C_{\rm III}$ are all taken clockwise. 
}
\label{fig:contours}
\end{figure}

This Appendix presents supplementary explanations about 
contours used for the integrals in the complex plane, which appear in Secs.~II and III. 
Figure \ref{fig:contours} shows the contours explained in this Appendix.

Here, we denote 
\begin{eqnarray}
\label{AZZ}
A(\zeta)
\equiv
\frac{Z(\zeta)}{1 + \kappa^{-2}
[ 1 + \zeta Z(\zeta) ]}
, 
\end{eqnarray}
and note that $A(\zeta)$ has no poles in the upper half-plane ${\rm Im}(\zeta)>0$. 
As shown in the second line of Eq.~(\ref{solE}), $A(\zeta) e^{-i \zeta \tau}$ is integrated 
along the real axis in the complex $\zeta$-plane from ${\rm Re}(\zeta)= - \infty$ to ${\rm Re}(\zeta)= + \infty$. 
We let $L_E$ denote the real axis as the contour for integration.  
We also consider $\tau$ to take a positive fixed value. 
Then, to apply the residue theorem for deriving the third line of Eq.~(\ref{solE}),  
we need to deform the integration contour from $L_E$ to the closed one by adding to $L_E$ the 
semicircle $C_-: \zeta = R e^{i \theta}$ in the lower half-plane ${\rm Im}(\zeta)<0$ 
where the radius is given by $R\gg 1$ and 
the orientation of $C_-$ is taken clockwise by varying the argument $\theta$ of $\zeta$
from $0$ to $-\pi$. 
We also choose $C_-$ such that $A(\zeta)$ has no poles on $C_-$. 
The deformation of the contour to $C_-$ mentioned above is justified because 
the part of the contour integral along the semicircle $C_-$ in the lower half-plane ${\rm Im}(\zeta)<0$
vanishes in the limit of $R \rightarrow + \infty$ as shown below.

Now, the semicircle $C_- : \zeta = R e^{i \theta}$ is 
divided into the three arcs $C_{\rm I}$, $C_{\rm II}$, and $C_{\rm III}$ which 
correspond to $0 > \theta >  - \pi / 4$, $- \pi / 4 > \theta > - 3 \pi / 4$, and 
$- 3\pi / 4 > \theta > - \pi$, respectively. 
We also denote the semicircle in the upper half-plane ${\rm Im}(\zeta)>0$ by 
$C_+ : \zeta = R e^{i \theta}$ where 
the orientation of $C_+$ is taken clockwise by varying the argument $\theta$ of $\zeta$
from $\pi$ to $0$. 
Then, $C= C_+ + C_-$ represents the circle with the radius $R \gg 1$ and the clockwise orientation. 
For $|\zeta | \equiv R \gg 1$, we have 
\begin{eqnarray}
\label{ZC}
Z (\zeta) \sim -  \zeta^{-1} 
&  \hspace*{3mm}  & 
\mbox{for $\zeta$ on $C_+$, $C_{\rm I}$, and $C_{\rm III}$}
,
\nonumber \\
| Z (\zeta) | \gg	 1
&  \hspace*{3mm} &
\mbox{for $\zeta$ on $C_{\rm II}$}
, 
\end{eqnarray}
from which we obtain 
\begin{equation}
\label{AZS}
A (\zeta) \sim 
\left\{
\begin{array}{ll}
- \zeta^{-1} 
&
\mbox{for $\zeta$ on $C_+$, $C_{\rm I}$, and $C_{\rm III}$}
\\
\kappa^2 \zeta^{-1} 
&
\mbox{for $\zeta$ on $C_{\rm II}$}
.
\end{array}
\right.
\end{equation}
Then, using Eq.~(\ref{AZS}), $\zeta  = R e^{i \theta} = R ( \cos \theta + i \sin \theta )$, 
$|d\zeta | = R | d\theta |$,  
and $\sin \theta \leq 2 \sqrt{2} \theta / \pi$ for $-\pi / 4 \leq \theta \leq 0$, 
we have 
\begin{eqnarray}
\label{ACI}
& & 
\left|
\int_{C_{\rm I}} d \zeta \; A(\zeta) e^{- i \zeta \tau}
\right|
 \leq 
\int_{C_{\rm I}} |d \zeta | \;  |A(\zeta) e^{- i \zeta \tau}|
\nonumber \\ 
&  & 
=
\int_{-\pi/4}^0 R d\theta \; 
|A( R e^{i \theta})| e^{\tau R \sin \theta}
\sim 
\int_{-\pi/4}^0 d\theta \; e^{ \tau R \sin \theta}
\nonumber \\ 
&  & 
\leq
\int_{-\pi/4}^0 d\theta \; e^{ 2\sqrt{2} \tau R \theta /\pi }
=
\frac{\pi}{2\sqrt{2} \tau R} 
\left( 
1 - e^{ - \tau R / \sqrt{2} }
\right)
. 
\end{eqnarray}
It can be shown from Eq.~(\ref{Zcc}) that 
\begin{equation}
\label{ACIII}
\int_{C_{\rm III}} d \zeta \; A(\zeta) e^{- i \zeta \tau}
=
\left( \int_{C_{\rm I}} d \zeta \; A(\zeta) e^{- i \zeta \tau} \right)^*
.
\end{equation}
Also, using Eq.~(\ref{AZS}) and $\sin \theta <  -1/\sqrt{2}$ for $-3\pi/4 \leq \theta \leq - \pi/4$, 
we obtain 
\begin{eqnarray}
\label{ACII}
& & 
\left|
\int_{C_{\rm II}} d \zeta \; A(\zeta) e^{- i \zeta \tau}
\right|
 \leq 
\int_{C_{\rm II}} |d \zeta | \;  |A(\zeta) e^{- i \zeta \tau}|
\nonumber \\ 
&  & 
=
\int_{-3 \pi/4}^{- \pi/4} R d\theta \; 
|A( R e^{i \theta})| e^{\tau R \sin \theta}
\sim 
\kappa^2
\int_{-3 \pi/4}^{- \pi/4} d\theta \; e^{ \tau R \sin \theta}
\nonumber \\ 
&  & 
\leq
\kappa^2
\int_{-3 \pi/4}^{- \pi/4} d\theta \; e^{ - \tau R /\sqrt{2}}
=
\frac{\pi}{2}
\kappa^2
 e^{ - \tau R /\sqrt{2} }
. 
\end{eqnarray}
We see from Eqs.~(\ref{ACI})--(\ref{ACII}) that,  for fixed $\tau >0$, 
the integrals of $A(\zeta) e^{- i \zeta \tau}$ along 
$C_{\rm I}$, $C_{\rm II}$, and $C_{\rm III}$ vanish in the limit of $R \rightarrow + \infty$. 
Since $C_- = C_{\rm I} + C_{\rm II} + C_{\rm III}$, we have 
\begin{equation}
\label{AC-}
\int_{C_-} d\zeta \;
A(\zeta) e^{- i \zeta \tau}
= 
0, 
\end{equation}
and accordingly 
\begin{equation}
\label{ALC-}
\int_{-\infty}^{+\infty} d\zeta \;
A(\zeta) e^{- i \zeta \tau}
= 
\int_{L_E} d\zeta \;
A(\zeta) e^{- i \zeta \tau}
= 
\int_{L_E + C_-} d\zeta \;
A(\zeta) e^{- i \zeta \tau}
, 
\end{equation}
in the limit of $R \rightarrow + \infty$. 
In Eq.~(\ref{ALC-}), $L _E + C_-$ is a closed integration contour, to which 
the residue theorem can be applied to derive the third line of Eq.~(\ref{solE}) as well as 
Eq.~(\ref{residue}). 
We can easily confirm that 
Eq.~(\ref{AC-}) is valid even though $A(\zeta)$ is replaced by 
$A(\zeta) / (\zeta - v/v_T)$, where  the value of $v/v_T$ is fixed when taking the large $R$ limit. 
Therefore, a closed integration contour including $C_-$ can also be used to 
apply the residue theorem in deriving Eq.~(\ref{solf1}).

Recalling that 
$A(\zeta) e^{- i \zeta \tau}$ has no pole on the upper half-plane $[{\rm Im}(\zeta) > 0]$,  
we can use Cauchy's integral theorem to replace the contour $L_E$ with $C_+$ in 
Eq.~(\ref{ALC-})  and obtain 
\begin{equation}
\label{ACC}
\int_{-\infty}^{+\infty} d\zeta \;
A(\zeta) e^{- i \zeta \tau}
= 
\int_{C_+} d\zeta \;
A(\zeta) e^{- i \zeta \tau}
= 
\int_{C} d\zeta \;
A(\zeta) e^{- i \zeta \tau}
, 
\end{equation}
where $C = C_+ + C_-$ is the closed circle with the radius $R \gg 1$. 
As shown in Eq.~(\ref{asymptotic2}), 
$A(\zeta)$ can be asymptotically expanded in $\zeta^{-1}$ 
for $\zeta \in C_+, C_{\rm I}, C_{\rm III}$ and $|\zeta|=R \gg 1$ 
as
\begin{equation}
\label{Aasymptotic}
A(\zeta)
=
- \frac{1}{\zeta}
\left[
1 + 
\sum_{n=1}^{N} e_n (\kappa^2) \zeta^{-2n}
+ {\cal O}(\zeta^{-2N-2})
\right]
. 
\end{equation}
Then, in the same way as in Eq.~(\ref{AC-}), 
we can show that the integration of the product of $e^{-i \zeta \tau}$ and 
the expression on the right-hand side of 
Eq.~(\ref{Aasymptotic}) over $C_- = C_{\rm I} + C_{\rm II} +C_{\rm III}$ 
vanishes in the large $R$ limit. 
Therefore, when we substitute Eq.~(\ref{Aasymptotic}) into Eq.~(\ref{ACC}), 
it is still valid even though the contour $C$ contains the arc $C_{\rm II}$ 
where Eq.~(\ref{Aasymptotic}) does not hold. 
Thus, we obtain the second line of Eq.~(\ref{asymptotic3}) where the residue theorem 
is also used to derive the third line. 
Taking the limit of $\tau \rightarrow +0$, Eqs.~(\ref{ACC}) and (\ref{asymptotic3}) give
\begin{equation}
\label{Ae0}
\lim_{\tau \rightarrow +0}
\int_{-\infty}^{+\infty} \frac{d\zeta}{2\pi} 
A(\zeta) e^{- i \zeta \tau}
= 
\lim_{\tau \rightarrow +0}
\int_{C} \frac{d\zeta}{2\pi}
A(\zeta) e^{- i \zeta \tau}
=
i
.
\end{equation}
Then, Eq.~(\ref{lim0int}) results from Eqs.~(\ref{Ae0}) and (\ref{residue}). 

Up to this point,  $\tau$ is assumed to be positive. 
We finally consider the case of $\tau = 0$ 
in which $e^{-i \zeta \tau} = 1$ 
and Eq.~(\ref{AC-}) does not hold. 
Again, using Eq.~(\ref{AZS}) in the large $R$ limit, we have 
\begin{eqnarray}
\label{A0}
\int_{C} \frac{d\zeta}{2\pi}
A(\zeta) 
& = & 
- 
\int_{C_+ + C_{\rm I} + C_{\rm III} } \frac{d\zeta}{2\pi}
\zeta^{-1}
+
\int_{C_{\rm II}} \frac{d\zeta}{2\pi}
\kappa^2 \zeta^{-1}
\nonumber \\
& = &  \frac{3}{4} i -  \frac{\kappa^2}{4}  i 
= \frac{i}{4} (3-\kappa^2)
,  
\end{eqnarray}
where the orientation of $C$ is taken clockwise. 
Then, Eq.~(\ref{A0}) is used to derive Eq.~(\ref{tau0int}) 
where the summation of the residues about the poles of $A(\zeta)$ 
are also shown. 


\color{black}
\section{Contour-dynamics simulation for Vlasov-Poisson system}

\label{sec:CD}
The Contour Dynamics (CD) method was proposed for solving inviscid and incompressible motions in fluid mechanics 
in the two-dimensional space.~\cite{Zabusky}.  
Here, we briefly describe the application of the CD method to the solution of the one-dimensional Vlasov-Poisson system 
with the periodic boundary condition [see Ref.~\cite{Sato} for details]. 

Normalizing the time $t$, spatial variable $x$, velocity $v$, distribution function $f(x, v, t)$, and electrostatic potential $\phi(x, t)$, 
appropriately, 
the Vlasov-Poisson equations can be written as 
\begin{equation}
\label{normaized_Vlasov}
\frac{ \partial f (x, v, t)}{\partial t } +v \frac{\partial f(x,v,t)}{\partial x} 
+ a(x, t) \frac{\partial f(x,v,t)}{\partial v} =0
,
\end{equation}
and
\begin{equation}
- \frac{ \partial \phi (x, t)}{\partial x^2 } = 1 - \int_{-\infty}^{+\infty} f(x, v, t) dv 
\equiv F(x, t)
,
\end{equation}
respectively. 
In Eq.~(\ref{normaized_Vlasov}), 
the acceleration field is given by 
$
a(x, t) \equiv \partial \phi (x, t) / \partial x
$.

In the CD method, $N_{max}$ contours $C_m(t)$ $(m=1, \cdots, N_{max})$ are considered in the $(x, v)$-plane.
Each point $(x, v)$ on $C_m(t)$ moves according to the equations of motion
\begin{equation}
\label{dxdtdvdt}
\frac{dx}{dt} = v, 
\hspace*{5mm}
\frac{dv}{dt} = a(x, t) 
.
\end{equation}
The solution $f(x, v, t)$ of the Vlasov equation is expressed as a piece-wise constant distribution function,
\begin{equation}
\label{f_pw}
f_{pw}(x, v, t) =  \sum_{m=1}^{N_{max}} \Delta f_m  \; I [(x, v) \in S_m (t) ] 
\end{equation}
where 
$I[P] = 1$ if $P$ is true and $I[P] = 0$ otherwise.
Here, $S_m (t)$ $(m=1, \cdots, N_{max})$ is the internal region of the $N_{max}$ contours $C_m(t)$ in the phase $(x, v)$ space at time $t$, and \( \Delta f_m \) denotes the jump of the distribution function across the contour \( C_m (t) \).
The electrostatic potential \( \phi (x, t) \)  
can be calculated from  
 \begin{equation}
 \label{potential}
   \phi(x, t) = \int_{-L/2}^{L/2} G(x; {\xi}) F({\xi}, t) d{\xi} + \text{const}
.
\end{equation}
Here, 
the Green function \( G \) is given by 
\begin{equation}
\label{Green1}
   G(x; \xi) = \frac{1}{2L}\left(|x - {\xi}| - \frac{L}{2}\right)^2
,
\end{equation}
which is the solution of 
\begin{equation}
\label{Green2}
\frac{ \partial G(x; \xi)}{\partial x^2 }   = \frac{1}{L} - \delta(x - {\xi})
\end{equation}
 with $L$ being the period length in the $x$ direction. 

     The acceleration \( a(x, t) \) of each particle is obtained using the CD representation as follows,
 \begin{equation}
  \label{ax}
 a(x) = \sum_{m} \Delta f_m \oint_{C_m} G(x; {\xi}) dv
.
\end{equation}
By substituting this into Eq.~(\ref{dxdtdvdt}) and integrating it in time, 
the time evolutions of the contours $C_m(t)$ are obtained, and the distribution function at each time $t$ is determined from Eq.~(\ref{f_pw}).
In practice, when performing numerical simulations, the contours $C_m(t)$ are divided into a finite number of nodes $(x_i, v_i)$. Between the nodes, the contour integral in Eq.~(\ref{ax}) is approximated by the integral along line segments, 
and the numerical solution of the equations of motion for the finite number of nodes $(x_i, v_i)$ is obtained.

In CD simulations, without using linear approximations, the solution of the nonlinear Vlasov equation is obtained in the form of 
the piece-wise constant distribution function as in Eq.~(\ref{f_pw}).
We use the piece-wise constant distribution function $f_{Mpw}(v)$ corresponding to the Maxwellian 
equilibrium velocity distribution, for which the outermost contours are placed at $v = \pm 5 v_t$, and $f_{Mpw}(v)=0$ for $|v| > 5 v_t$.
For all CD simulations in this study, 
we follow the same procedure as in Ref.~\cite{Sato} to give $f_{pw}(x, v, t=0)$ that 
corresponds to the initial condition in Eq.~(\ref{initial}). 
The number of contours used is 200, with each contour represented by 100 nodes.
At time $t$, the value of the distribution function $f(x, v, t)$ at any point $(x, v)$ in the phase space 
is obtained by the linear Hermite interpolation from the values of the distribution function on the contour 
to which the nodes $(x_i, v_i)$ in the vicinity of $(x, v)$ belong, and is used for comparison 
with theoretical predictions.

\section{Time reversibility of Vlasov-Poisson System}

In the Vlasov-Poisson system, 
the time reversal map $T$ from the two-dimensional phase space to itself 
is defined by 
\begin{equation}
T (x, v) = (x, -v)
,
\end{equation}
which satisfies 
\begin{equation}
T^2 \equiv T \circ T = I
,
\end{equation}
where $I$ represents the identity map. 
We can easily confirmed that, 
when $f(x, v, t)$ is a solution of the Vlasov equation, 
$f(T(x, v), -t) = f(x, -v, -t)$ is a solution as well. 
It should be noted that 
the property of the time reversal map mentioned above 
holds whether the linear or nonlinear case is considered. 
When the initial condition in Eq.~(\ref{initial}) is employed, 
the initial distribution function 
$f (x, v, t=0)$ satisfies 
the condition 
$f(x, - v, t=0) = f(x, v, t=0)$.
Then, for the solution $f(x, v, t)$ under this initial condition, 
$f(x, -v, -t)$ becomes the solution satisfying the same initial condition. 
Thus, from the uniqueness of the solution under the same initial condition,   
we find that $f(x, -v, -t) = f(x, v, t)$ holds for any time $t$ and that 
$E(x, -t) = E(x, t)$ results from Poisson's equation.

\section{Information entropies and mutual information in Vlasov-Poisson system}

We here regard the electron's position and velocity as random variables denoted 
by $X$ and $V$. 
The joint probability density function of $(X, V)$ is represented by 
$p(x, v)$ which satisfies the normalization condition, 
\begin{equation}
 \int_{-L/2}^{+L/2} dx \int_{-\infty}^{+\infty} dv \; p(x, v) = 1
.
\end{equation}
It is related to the distribution function in Eq.~(\ref{Vlasov1}) by 
\begin{equation}
\label{pxv}
p(x, v) = \frac{1}{n_0 L} f(x, v)
,
\end{equation}
where $n_0$ is the average number density of electrons. 
Here, the functions $p$ and $f$ depends on time $t$ although 
$t$ is omitted from the arguments of these functions for simplicity. 
The marginal probability distribution functions for $X$ and $V$ are 
defined by 
\begin{equation}
\label{px}
p_X(x) = \int_{-\infty}^{+\infty} dv \; p(x, v) =\frac{1}{L} \frac{n(x)}{n_0}
\end{equation}
and 
\begin{equation}
\label{pv}
p_V(v) =  \int_{-L/2}^{+L/2} dx \; p(x, v) =\frac{1}{n_0} \langle f \rangle (v)
, 
\end{equation}
respectively. 
Here, the electron density is denoted by $n(x)$. 

The information (or Shannon) entropy~\cite{Information} is originally defined for probability distributions 
of discrete variables. 
Rigorously speaking, 
for probability distributions of continuous variables, 
it should be called the differential entropy.
When dividing the interval $[-L/2, L/2)$ of the variable $x$ is divided into 
$N_x$ intervals of equal width, 
the central value  $x_i$ in each interval is given by
\begin{equation}
x_i = i \Delta x , \hspace*{5mm}
\Delta x = \frac{L}{N_x}
, \hspace*{5mm}
i = - \frac{(N_x-1)}{2}, \cdots, \frac{(N_x-1)}{2}
.
\end{equation}
In the same way,  when dividing the interval $[-v_{\rm max}, v_{\rm max}]$ 
of the variable $v$ into $N_v$ intervals of equal width, 
the central value $v_j$ in each interval is given by 
\begin{equation}
v_j = j \Delta v , \hspace*{5mm}
\Delta v = \frac{2 v_{\rm max}}{N_v}
, \hspace*{5mm}
i = - \frac{(N_v-1)}{2}, \cdots, \frac{(N_v -1)}{2}
.
\end{equation}
Setting 
\begin{equation}
z_{ij} = (x_i, v_j) , \hspace*{5mm}
N_z = N_x N_v 
,
\end{equation}
and letting $v_{\rm max}$ be sufficiently large,
we consider the $(x, v)$ space as a collection of $N_z$ cells, 
each of which is centered at $z_{ij} = (x_i, v_j)$. 
Then, we approximate continuous variables $(x, v)$ by discrete ones 
$z_{ij} = (x_i, v_j)$, and relate $p(x, v)$ to the probability distribution function $P(x_i, v_j)$ 
of the discrete variables $(x_i, v_j)$ by
\begin{equation}
p( x, v ) \Delta x \Delta v=  P(x_i, v_j) 
,
\end{equation}
which satisfies the normalization condition,
\begin{equation}
 \int_{-L/2}^{+L/2} dx \int_{-\infty}^{+\infty} dv \; p(x, v) = 
 \sum_{i,j}  P(x_i, v_j) = 1
 .
\end{equation}
Using $P(x_i, v_j)$, 
the information entropy is defined by 
\begin{equation}
S[P] 
\equiv
 \sum_{i,j} P(x_i, v_j) S(x_i, v_j)
\equiv
- \sum_{i,j} P(x_i, v_j) \log P(x_i, v_j)
,
\end{equation}
where 
\begin{equation}
S(x_i, v_j)
\equiv
- \log P(x_i, v_j)
\end{equation}
represents the self-entropy which is also called 
the self-information or information content. 
Similarly, we employ $p(x, v)$ to define the differential entropy by
\begin{eqnarray}
S[p] 
& \equiv &
 \int_{-L/2}^{+L/2} dx \int_{-\infty}^{+\infty} dv \; p(x, v)  s(x, v)
 \nonumber \\
& \equiv &
- \int_{-L/2}^{+L/2} dx \int_{-\infty}^{+\infty} dv \; p(x, v) \log p(x, v)
\end{eqnarray}
where 
\begin{equation}
s(x, v)
\equiv
- \log p(x, v)
.
\end{equation}
We can easily derive the following relations,
\begin{equation}
S(x_i, v_j)
\equiv
s(x, v)
- \log (\Delta x \Delta v)
,
\end{equation}
and
\begin{equation}
\label{SPSp}
S[P] 
\equiv
S[p]
- \log (\Delta x \Delta v)
.
\end{equation}
Here, we should carefully note that 
$S(x_i, v_j)$ and $S[P]$ are non-negative while 
$s(x, v)$ and $s[p]$ can take both positive and negative values but satisfy 
the conditions,  
\begin{equation}
s(x, v)
\geq \log (\Delta x \Delta v)
, \hspace*{5mm}
S[p] 
\geq \log (\Delta x \Delta v)
.
\end{equation}

From the marginal probability distribution functions 
$p_X(x)$ and $p_V(v)$, 
we can define the marginal distribution functions $P_X(x_i)$ and $P_V(v_j)$ 
of the discrete variables 
$x_i$ and $v_j$ by 
\begin{equation}
P_X(x_i)
= 
\sum_j P(x_i, v_j)
, 
\hspace*{5mm}
P_v(v_j)
= 
\sum_i P(x_i, v_j)
,
\end{equation}
which are related to $p_X(x)$ and $p_V(v)$ by 
\begin{equation}
P_X(x_i)
= 
p_X (x) \Delta x
, 
\hspace*{5mm}
P_V(v_j)
= 
p_V (v) \Delta v
.
\end{equation}
Then, 
$P_X(x_i)$ and $P_V(v_j)$ 
are used to define the information entropies 
$S[P_X]$ and $S[P_V]$ as 
\begin{eqnarray}
S[P_X] 
& \equiv & 
 \sum_{i} P_X(x_i) S_X(x_i)
\equiv
- \sum_{i} P_X(x_i) \log P_X(x_i)
\nonumber \\
S[P_V] 
& \equiv & 
 \sum_{i} P_V(v_j) S_V(v_j)
\equiv
- \sum_{i} P_V(v_j) \log P_V(v_j)
,
\hspace*{5mm}
\end{eqnarray}
where the self-entropies $S_X(x_i)$ and $S_V(v_j)$ 
are defined by 
\begin{equation}
\label{SEXV}
S_X(x_i)
\equiv
- \log P_X(x_i),
\hspace*{5mm}
S_V(v_j)
\equiv
- \log P_V(v_j)
.
\end{equation}
Similarly, we use $p_X(x)$ and $p_V(v)$ to 
the entropies $S[p_X]$ and $S[p_V]$ by 
\begin{eqnarray}
S[p_X] 
& \equiv &
 \int_{-L/2}^{+L/2} dx\; p_X(x)  s_X(x)
 \equiv 
 - \int_{-L/2}^{+L/2} dx \; p_X(x) \log p_X(x)
 \nonumber \\
 S[p_V] 
& \equiv &
 \int_{-\infty}^{+\infty} dv\; p_V(v)  s_V(v)
 \equiv 
 -  \int_{-\infty}^{+\infty} dv\; p_V(v)  \log p_V(v)
 ,
 \nonumber \\ & & 
\end{eqnarray}
where
\begin{equation}
s_X(x)
\equiv
- \log p_X(x)
,
\hspace*{5mm}
s_V(v)
\equiv
- \log p_V(v)
.
\end{equation}
We can also confirm the following relations,
\begin{eqnarray}
S_X(x_i)
& \equiv &
s_X(x)
- \log (\Delta x)
\geq 0
\nonumber \\
S_V(v_j)
& \equiv &
s_V(v)
- \log (\Delta v)
\geq 0
\end{eqnarray}
and
\begin{eqnarray}
\label{SPSpXV}
S[P_X] 
& \equiv & 
S[p_X]
- \log (\Delta x)
\geq 0
\nonumber \\
S[P_V] 
& \equiv & 
S[p_V]
- \log (\Delta v)
\geq 0
.
\end{eqnarray}
The conditional entropies $S_P(V|X)$ 
and $S_P(X|V)$ are defined from 
the probability distribution functions $P(x_i, v_j)$, $P_X(x_i)$, and $P(v_j)$ by 
\begin{eqnarray}
& & 
S_P(V|X)
 \equiv  
- \sum_{i, j} P (x_i, v_j )
 \log \left(
 \frac{P(x_i, v_j)}{P_X(x_i)}
\right)
\nonumber \\
& & 
= S[P] - S[P_X]
=S_P(X,V) - S_P(X)
\geq 0
,
\nonumber \\
& & 
S_P(X|V)
 \equiv  
- \sum_{i, j} P (x_i, v_j )
 \log \left(
 \frac{P(x_i, v_j)}{P(v_j)}
\right)
\nonumber \\
&  & 
= S[P] - S[P_V]
=S_P(X,V) - S_P(V)
\geq 0
. 
\end{eqnarray}
Similarly, the conditional entropies $S_p(V|X)$ and 
$S_p(X|V)$ are defined from $p (x, v )$, $p_X(x)$, 
and $p_V(v)$ by 
\begin{eqnarray}
\label{cond_entropies}
& & 
S_p(V|X)
 \equiv  
- \int_{-L/2}^{L/2} dx \int_{–\infty}^{+\infty} dv \; 
p (x, v )
 \log \left(
 \frac{p(x, v)}{p_X(x)}
\right)
\nonumber \\
&  & 
= S[p] - S[p_X]
= S_p(X,V) - S_p(X)
\geq \log (\Delta v)
,
\nonumber \\
& & 
S_p(X|V)
 \equiv 
- \int_{-L/2}^{L/2} dx \int_{–\infty}^{+\infty} dv \; 
p (x, v )
 \log \left(
 \frac{p(x, v)}{p_V(v)}
\right)
\nonumber \\
& & 
= S[p] - S[p_V]
= S_p(X,V) - S_p(V)
\geq \log (\Delta x)
.
\end{eqnarray}
To represent the mutual dependence of the random variables 
$X$ and $V$, 
the mutual information $I (X, V)$
is defined by 
\begin{eqnarray}
& & 
I (X, V)
 \equiv  
- \sum_{i, j} P (x_i, v_j )
 \log \left(
 \frac{P(x_i, v_j)}{P_X(x_i)P_V(v_j)}
\right)
\nonumber \\
& & 
= S_P(X) + S_P(V) - S_P (X, V) 
\nonumber \\
& & 
= S_P(X) - S_P(X|V) 
= S_P(V) - S_P(V|X) 
\nonumber \\
& & 
= 
- \int_{-L/2}^{L/2} dx \int_{–\infty}^{+\infty} dv \; 
p(x, v) 
 \log \left(
 \frac{P(x, v)}{P_X(x)P_V(v)}
\right)
\nonumber \\
& & 
= S_p(X) + S_p(Y) - S_p (X, Y) 
\nonumber \\
& & 
= S_p(X) - S_p(X|V) 
= S_p(V) - S_p(V|X) 
\geq 0
.
\end{eqnarray}
The mutual information takes the same non-negative value whether it is defined from 
the distribution functions of the discrete variables or from those of the continuous variables.

Hereafter, we express the time dependence of the distribution function by writing 
$
p(x, v, t) = f(x, v, t) / (n_0 L)
$, 
$
p_X(x, t) = \int_{-\infty}^{+\infty} dv \; p(x, v, t) = n(x, t)/(n_0 L)
$,
and
$
p_V(v, t) =  \int_{-L/2}^{+L/2} dx \; p(x, v, t) = \langle f \rangle (v, t) / n_0
$.
It is well known that, for the solution $f(x, v, t)$ of the Vlasov equation in Eq.~(\ref{Vlasov1}), 
the entropy $S [p]$ defined from $p(x, v, t) = f(x, v, t) / (n_0 L)$ by 
\begin{equation}
S [p] \equiv S_p(X, V) 
\equiv  - \int_{-L/2}^{+L/2} dx  \int_{-\infty}^{+\infty} dv \; p(x, v, t) \log p(x, v, t)
\end{equation}
is an invariant, 
\begin{equation}
\frac{d}{dt} S [p] \equiv \
\frac{d}{dt}  S_p(X, V) 
=0
. 
\end{equation}
On the other hand, 
the entropies defined from $p_X(x, t)$ and $p_V(v, t)$ by
\begin{equation}
S [p_X] \equiv S_p(X) 
\equiv - \int_{-L/2}^{+L/2} dx  \; p_X(x, t) \log p_X(x, t)
\end{equation}
and
\begin{equation}
S [p_V] \equiv S_p(V) 
\equiv  -  \int_{-\infty}^{+\infty} dv \; p_V(v, t) \log p_V(v, t)
\end{equation}
depend on time $t$ in general. 

The relative entropy (or Kullback-Leibler divergence) of the distribution $p_V(v, t)$ at time $t$ 
relative to the initial distribution $p_V(t,0)$ is defined as
\begin{equation}
\label{relative_entropy}
S (p_V, t || p_V, 0)
\equiv  
 \int_{-\infty}^{+\infty} dv  \; p_V(v, t) 
 \log \left(
 \frac{p_V(v, t)}{p_V(v, 0)}
\right)
,
\end{equation}
which takes only non-negative values and vanishes 
if and only if $p_V(v, t) = p_V(v, 0)$. 
The non-negativity is derived from the inequality $\log (x^{-1}) = - \log x \geq 1 - x$ for $x  > 1$.
The relative entropy $S (p_V, t || p_V, 0)$ is the average difference 
between the information quantities, 
$s_V(v, 0) - s_V(v, t) = - \log p_V(v, 0) - ( - \log p_V(v, t) ) =
\log [p_V(v, t) / p_V(v, 0) ]$,  and represents the amount of information lost 
when approximating the correct velocity distribution $p_V(v, t)$ at time $t$ by the initial distribution function $p_V(v, 0)$.
From Eq.~(\ref{fut2}), we have 
\begin{equation}
\Omega_t(v_0) 
= \log
\left[
\frac{p_V (u(v_0, t), t)}{p_V (u(v_0, t),  0)}
\right]
,
\end{equation}
and 
\begin{eqnarray}
\label{relative_entropy2}
& & 
S (p_V, t || p_V, 0)
= 
 \int_{-\infty}^{+\infty} dv_0  
\; p_V (v_0, 0)
\log 
\biggl[ 
\frac{p_V (u(v_0, t), t)}{p_V (u(v_0, t), 0)}
\biggr]
\nonumber \\ & & 
=
 \int_{-\infty}^{+\infty} dv_0  
\; p_V(v_0, 0)
\;
 \Omega_t (v_0) 
 .
\end{eqnarray}
When the initial distribution function is the Maxwellian, 
using Eqs.~(\ref{fexpOmega}) and (\ref{relative_entropy2}) leads to 
\begin{equation}
\label{SDEDS}
S (p_V, t || p_V, 0)
= 
\frac{\Delta {\cal E}(t)}{n_0 T} - \Delta S[P_X](t)
\geq 0
. 
\end{equation}
Here,  $\Delta {\cal E}(t)$ is the increase in the kinetic energy of electrons per unit volume 
and equals the decrease in the electric field energy density, 
\begin{eqnarray}
\label{DEDQ}
\Delta {\cal E}(t) & = & 
\int_{-\infty}^{+\infty} dv \; f_0 (v_0) \Delta {\cal E}(u(v_0, t), t)
\nonumber \\ 
& = &  
 - \frac{\langle E^2 \rangle (t) - \langle E^2 \rangle (0)}{8\pi}
 \nonumber \\
 & = & 
 - n_0 \Delta Q(X|V)
. 
\end{eqnarray}
We now call two systems described by $p_X(x)$ and $p_V(v)$ 
as the $X$-system and the $V$-system, respectively. 
Then, we can regard
$\Delta Q(X|V) = - \Delta {\cal E} / n_0$ in Eq.~(\ref{DEDQ})
as the energy transfer per electron from the $V$-system to the $X$-system. 
Recalling that $\Delta S[P_X](t)$ represents the increase in 
the entropy $S[P_X]$ during the time interval from 0 to $t$
and that $S_P(X, V)$ is time-independent, 
we can use Eq.~(\ref{cond_entropies}) to get  
\begin{equation}
\label{DSPV2}
 \Delta S[P_V] = \Delta S_P (V)  =  - \Delta S_P(X|V)
 .
\end{equation}
From Eqs.~(\ref{SDEDS}), (\ref{DEDQ}), and (\ref{DSPV2}), 
we find 
\begin{equation}
\label{2ndlaw}
S (p_V, t || p_V, 0)
= 
\Delta S_P(X|V) - \frac{\Delta Q(X|V)}{T}
\geq 0
 .
\end{equation}
Since the initial distribution $p_V(v, 0)$ is given by the Maxwellian,   
we regard the $V$-system here as the thermal reservoir with the 
temperature $T$. 
Then, Eq.~(\ref{2ndlaw}) implies that 
the heat transfer $\Delta Q(X|V)$ from the thermal reservoir 
to the $X$-system and the change in the conditional 
entropy of the $X$-system in contact with the thermal reservoir 
satisfy the inequality in the form of the second law of thermodynamics.

When the values of $p_V (v, t)$ and $p_V (v,0)$ are close to each other, 
we obtain 
\begin{eqnarray}
\label{relative_entropy3}
& & S (p_V, t || p_V, 0) 
=
-
\int_{-\infty}^{+\infty} dv \;
p_V  (v, t) 
\log \left[
\frac{ p_V  (v, 0)}{
p_V  (v, t) 
}
\right]
\nonumber \\
&  &  \simeq
\int_{-\infty}^{+\infty} dv \;
p_V  (v, t) 
\left[
-
\left( 
\frac{ p_V  (v, 0)}{
p_V  (v, t) 
}
 - 1 \right)
+ \frac{1}{2}
\left(
\frac{ p_V  (v, 0)}{
p_V  (v, t) 
}
- 1 \right)^2 
\right]
\nonumber \\
&  &  =
\frac{1}{2}
\int_{-\infty}^{+\infty} dv \;
p_V  (v, t) 
\left(
\frac{p_V  (v, 0)}{
p_V  (v, t) 
}
- 1 \right)^2 
\nonumber \\ & & 
=
\frac{1}{2}
\int_{-\infty}^{+\infty} dv \;
\frac{ [ p_V  (v, t) - p_V  (v, 0)  ]^2}{
p_V  (v, t) 
}
\nonumber \\ & & 
\simeq
\frac{1}{2}
\int_{-\infty}^{+\infty} dv \;
\frac{ [ p_V  (v, t) - p_V  (v, 0)  ]^2}{
p_V  (v, 0) }
.
\end{eqnarray}
Using Eq.~(\ref{relative_entropy3}) and 
the ordering parameter $\alpha$ for the perturbation amplitude,  
$S (p_V, t || p_V, 0) = {\cal O}(\alpha^4)$ is derived.
We also find from Eqs.~(\ref{SDEDS}) and (\ref{DSPV2}) that
$\Delta {\cal E}(t)/(n_0 T) = \Delta S[P_X](t)$ and 
$\Delta S_P(X|V) = \Delta Q(X|V)/T$ 
hold up to ${\cal O}(\alpha^2)$.

\nocite{*}
\bibliography{aipsamp}

\end{document}